# Reconstructing large networks with time-varying interactions

**Authors:** Chun-Wei Chang[1,2], Takeshi Miki[3,4,5], Masayuki Ushio[6,7], Hsiao-Pei Lu[8], Fuh-Kwo Shiah[2,3], Chih-hao Hsieh[1,2,3,9*]

**Affiliations:**
[1]National Center for Theoretical Sciences, Taipei 10617, Taiwan
[2]Research Center for Environmental Changes, Academia Sinica, Taipei 11529, Taiwan
[3]Institute of Oceanography, National Taiwan University, Taipei 10617, Taiwan
[4]Department of Environmental Engineering and Ecology, Faculty of Advanced Science and Technology, Ryukoku University, Otsu, Shiga, 520-2194, Japan
[5]Center for Biodiversity Science, Ryukoku University, Otsu, Shiga, 520-2194, Japan
[6]Hakubi Center, Kyoto University, Kyoto, Kyoto 606-8501, Japan
[7]Center for Ecological Research, Kyoto University, Otsu, Shiga 520-2113, Japan
[8]Department of Biotechnology and Bioindustry Sciences, National Cheng Kung University, Tainan, Taiwan
[9]Institute of Ecology and Evolutionary Biology, Department of Life Science, National Taiwan University, Taipei 10617, Taiwan

\* Correspondence to: Chih-hao Hsieh; **Email:** chsieh@ntu.edu.tw

**Number of words in abstract: 196**

**Number of words in main text: 4252**

**Number of cited references: 47**

**Number of tables & figures: 4 figures and 1 table**

**Abstract**

Reconstructing interactions from observational data is a critical need for investigating natural biological networks, wherein network dimensionality (i.e. number of interacting components) is usually high and interactions are time-varying. These pose a challenge to existing methods that can quantify only small interaction networks or assume static interactions under steady state. Here, we proposed a novel approach to reconstruct high-dimensional, time-varying interaction networks using empirical time series. This method, named "multiview distance regularized S-map", generalized the state space reconstruction to accommodate high dimensionality and overcome difficulties in quantifying massive interactions with limited data. When we evaluated this method using the time series generated from a large theoretical model involving hundreds of interacting species, estimated interaction strengths were in good agreement with theoretical expectations. As a result, reconstructed networks preserved important topological properties, such as centrality, strength distribution and derived stability measures. Moreover, our method effectively forecasted the dynamic behavior of network nodes. Applying this method to a natural bacterial community helped identify keystone species from the interaction network and revealed the mechanisms governing the dynamical stability of bacterial community. Our method overcame the challenge of high dimensionality and disentangled complex time-varying interactions in large natural dynamical systems.

**Introduction**

Interaction network critically determines the dynamics and stability of dynamical systems (Albert *et al.* 2000; Strogatz 2001). Therefore, statistical tools for analyzing network properties have been well developed (Barrat *et al.* 2004). However, quantitative recovery of interaction networks from natural systems remains challenging, in particular, to identify and quantify network edges (i.e., interactions between network nodes). For instance, identifying trophic interactions in food webs relies on direct observations, e.g., gut content analysis (Morinière *et al.* 2003) or indirect molecular approaches, e.g., stable isotope analysis (Fry 2006). These approaches are labor-intensive and have limited resolution to identify and quantify interactions in large networks. Other types of interactions, such as competition and cooperation, are ambiguously defined in real-world situations and can be only determined for a limited number of interacting components (Griffin *et al.* 2004). Consequently, it becomes extremely difficult to reconstruct high-dimensional interaction networks consisting of enormous numbers of nodes (e.g., microbes living in natural ecosystems and proteins in cells) and edges (e.g., competition/facilitation of microbial species and activation/inactivation of proteins).

An even more challenging aspect is that most of real-world interactions are not static but vary with states (or contexts) of dynamical systems (Deyle *et al.* 2016; Song *et al.* 2020). For instance, strength of competitive interactions among ecological populations depends on states of environmental variables (Deyle *et al.* 2016) or predators (Rogers *et al.* 2020); in any case, these states are time-varying (Rogers *et al.* 2020). Consequently, network properties (e.g., mean and standard deviation of interaction strengths) are also expected to be dynamic and cannot be recovered from conventional network reconstruction methods relying on steady state assumptions (Berry & Widder 2014; Xiao *et al.* 2017). Because network properties reflect the resilience and robustness of systems (Montoya *et al.* 2006), investigating dynamical behaviors of network properties is beneficial for predicting responses of biological systems to external disturbances. For instance, it remains unclear how food webs would respond to climate changes (Hoegh-Guldberg & Bruno 2010), or how cancer protein-protein interaction networks would respond to chemotherapy (Zinzalla & Thurston 2009).

More recently, empirical dynamic modeling (EDM) based on state space reconstruction (SSR) of dynamical attractor was proposed to quantify interactions in nonlinear dynamical systems. Convergent cross mapping (CCM) (Sugihara *et al.* 2012) detects causal interactions, whereas multivariate S-map (Deyle *et al.* 2016) quantifies time-varying interaction strengths. Although CCM can determine pairwise causal relationships, allowing to reconstruct high-dimensional causal networks, this approach cannot quantify the sign and time-varying strength of interaction. Conversely, S-map quantifies both the sign and dynamics of interactions, but it allows only a limited number of variables (or network nodes) to be embedded into an S-map model (Yu *et al.* 2020). In that regard, as an SSR-based method, S-map needs to be operated at the optimal embedding dimension (Deyle *et al.* 2016; Ushio *et al.* 2018; Cenci *et al.* 2019) and that is usually much smaller than the real number of interacting nodes (i.e., network dimensionality denoted as $m$). This restriction arises due to the "curse of dimensionality" (Bellman 1957); high dimensionality likely makes data too sparse to correctly depict neighboring relationships among data points (Hastie *et al.* 2009), whereas a correct neighborhood relationship is essential for applying SSR-based methods (Ye & Sugihara 2016). Although a modified S-map approach (Cenci *et al.* 2019) incorporating regularization (Hastie *et al.* 2009) may be helpful for estimating high-dimensional parameters, this method is still subject to the curse of dimensionality in SSR and only allows embedding much fewer variables than the true number of interacting nodes in real-world networks (Ushio *et al.* 2018; Yu *et al.* 2020). Consequently, an appropriate method to reconstruct high-dimensional, time-varying interaction networks for natural biological systems is still lacking.

Here, we propose a novel approach to reconstruct high-dimensional (i.e., large $m$), time-varying interaction networks in large, nonlinear dynamical systems (e.g., ecological communities). The advance of our approach is developing a novel distance measure (Fig. 1) that quantifies the neighboring relationships among high-dimensional data points in SSR. We refer to this measure as "multiview distance" because it is determined by ensembling numerous distances measured in various low-dimensional, topologically equivalent SSRs (i.e., multiview embeddings (Ye & Sugihara 2016)). Through multiview distance, our approach links two existing EDM methods, multiview embedding (Ye & Sugihara 2016) and regularized S-map (Cenci *et al.* 2019), to reconstruct large interaction networks, but avoids problems caused by high dimensionality when conducting SSR operations under the optimal embedding dimension (*E*) much smaller than network dimensionality (i.e., $E \ll m$). This method, named "multiview distance regularized S-map" (abbreviated as MDR S-map, hereafter), is summarized in two steps (Fig. 1): (i) measuring multiview distances among *m*-dimensional data from various *E*-dimensional multiview SSR that embeds various combinations of variables (Ye & Sugihara 2016); and (ii) determining *m*-dimensional interaction matrix at each time point by plug-in multiview distances to S-map algorithm with regularization constraints (Cenci *et al.* 2019) (Methods and SI Text, *I. MDR S-map algorithm and implementation*). More detail statistical properties of MDR-S-map were also offered in SI Text, *II. Statistical properties of MDR S-map.* Through the two-step procedure of MDR S-map, we reconstructed time-varying, high-dimensional interaction networks for large dynamical systems using time series data.

We tested the MDR S-map method using simulated time-series data from a stochastic population dynamic model (*m*=117 variables and *N*=100 time points; Methods and SI Text, *III. Parameterization of theoretical models*), where interaction strengths and network properties are exactly known. In particular, we examined whether the reconstructed networks preserved the three types of network properties (Methods), including: i) topological properties of network node; ii) topological properties of the entire network; and iii) network stability measures. Then, we employed this method to analyze empirical bacterial community data (dominant OTUs (=operational taxonomic units) with *m*=136 and *N*=90) collected from a natural coastal environment (Martin-Platero *et al.* 2018). In both simulated and empirical datasets, numbers of selected variables (i.e., *m*=117 and 136 for simulated and empirical bacterial community data, respectively) were more than the length of time series data (*N*=100 and 90 for simulated and empirical data, respectively), a situation that very likely occurs in real-word datasets concerning large interaction networks. Based on analysis of a bacterial community, we demonstrated a real-world application of MDR S-map for reconstructing large, time-varying interaction networks and unveiling the interplay among network properties, dynamics and stability.

## Materials and Methods

*MDR S-map*

Our method of reconstructing large interaction networks using time series data was based on EDM (an approach rooted in attractor reconstruction). For attractor reconstruction, a critical parameter is the optimal embedding dimension (Kennel *et al.* 1992; Hsieh *et al.* 2005; Shalizi 2006; Deyle *et al.* 2016; Ye & Sugihara 2016). In large systems, we encountered an issue: the optimal embedding dimension was often much smaller than the number of interacting components (network nodes); in other words, the number of interacting components was much greater than the optimal embedding dimension (usually <20 (Hsieh *et al.* 2005)) can accommodate. To overcome this difficulty, we proposed a method, MDR S-map, that allows estimating interaction strengths of high-dimensional systems while maintaining the SSR operation at the low-dimensional, optimal embedding dimension. Figure 1 summarized the two-step analytical procedure based on state space reconstruction and more detailed algorithms, implementation, and statistical properties of MDR S-map were provided in SI Text, *I. MDR S-map algorithm and implementation* and II. *Statistical properties of MDR S-map*.

*Assessing MDR S-map based on a theoretical network model*

We validated the MDR S-map method using simulated time series data generated from a theoretical network model, as that interaction strengths and network topological properties are known *a priori*. This network model initially consists of 1000 mutually interacting species. Let $N_i(k)$ denote the population size of species $i$ ($i$ = 1, 2, …, 1000) at time step $t$ ($t$ = 0, 1, 2, …, $n$), and $\mathbf{N}(t) = (N_1(t), N_2(t), …, N_{1000}(t))^T$. Population dynamics of species $i$ followed the multi-species Ricker model,

$$N_i(t+1) = N_i(k)\exp\left[r_{0i}\left(1 + e_i\mathbf{MN}(t)\right)\right] \text{------(Eq. 3.)},$$

where $r_{0i}$ is intrinsic growth rate of species $i$, $e_i$ is a vector taking zero values for all except for the *i*th entity, **M** represents time-independent interaction matrix with the size 1000 x 1000 ($m_{ij}$ represent the effect of species *j* on species *i*). The detail parameterization of the multi-species Ricker model was offered in SI Text, *III. Parameterization of theoretical models*.

*Analyzing model interaction networks*

For model time series, we only used the last 100 time steps out of the 1000 steps simulations for further analyses. Because the model populations had chaotic dynamics and only some of these populations could co-exist until the end of model computations, we only selected dominant species with relative abundance > 0.1% (calculated over the last 100 steps) for further analyses. These dominant species had obvious temporal fluctuations, which is necessary for applying EDM. In total, 117 species were selected. Then, we applied the MDR S-map to reconstruct the interaction networks from the model time series and compared the reconstructed networks with the theoretical networks derived from the Jacobian matrix of the difference equation models. To make reasonable comparisons, we multiplied the theoretical expectation of Jacobian ($\frac{\partial X_{t+1}}{\partial Y_t}$) by the ratio of standard deviations, $\frac{\sigma_Y}{\sigma_X}$, because time series data were normalized with respect to the temporal standard deviations prior to performing MDR S-map. Finally, we examined whether the estimated interaction strengths were consistent with the scaled theoretical Jacobians.

Based on the reconstructed interaction networks, we aimed to verify whether or not: 1) the interaction strengths can be successfully recovered at each time point; and 2) topological properties of network can be preserved. In particular, we focused on three types of topological properties quantified in directed and weighted networks. First, we considered the topological properties that evaluated the importance of each network node, including in-, out-, and all interaction strengths (Barrat *et al.* 2004). We also computed cumulative distributions of interaction strengths (Emmerson & Yearsley 2004) and tested whether these reconstructed distributions preserved the shape of theoretical distributions. In addition, we considered more complex indices, including eigenvector centrality (Bonacich 1987), hub (Kleinberg 1999), and authority (Kolaczyk & Csárdi 2014), to evaluate the importance of each network node according to how they interacted with other nodes and where they located in the interaction networks. Second, we computed topological indices summarizing the entire interaction network, including mean transitivity (Barrat *et al.* 2004), as well as mean and standard deviation of interaction strengths. Finally, we evaluated the network stability based on instability indices, including local Lyapunov instability (Ushio *et al.* 2018) and structural instability (Cenci & Saavedra 2019), computed from the absolute values of dominant eigenvalues and traces (i.e., sum of the diagonal elements) of Jacobian matrices, respectively. The aim of stability analysis is to verify whether our reconstructed interaction networks can help evaluate the stability of large dynamical systems. In summary, we derived all three types of topological indices in order to test whether topological properties in theoretical networks were preserved in the reconstructed networks.

We also tested robustness of the MDR S-map on three data issues that are common in real world applications using the model data: 1) only percentage data are available; 2) impacts of data noise (including process and measurement error); 3) not all data from every network node are available (incomplete network nodes). The details are explained in SI Text.

*Analysis of empirical bacterial time series data*

We applied MDR S-map on empirical time series data of natural bacterial communities in Canoe Beach, Boston, MA, USA (Martin-Platero *et al.* 2018). This dataset was derived from 16S rRNA gene amplicon-sequencing data sampled every day between July 23, 2010 and October 23, 2010. In total, there were 90 valid time points, apart from 3 interrupted missing data. In this dataset, only relative abundances were available for bacteria OTUs. To be consistent with analysis of model time series, we selected the dominant species (>0.1% relative abundance). In total, 136 OTUs were selected. Due to data limitations, we reconstructed bacterial interaction networks using relative abundance data; nevertheless, based on our analysis using the model example, the network reconstruction using percentage data was still reliable to a large extent (See our discussion in SI Text, *II-2. Statistical properties of MDR S-map based on percentage data*)).

The reconstructed interaction networks enabled us to estimate how stability of the bacteria community changed through time and why. To investigate causal mechanisms underlying structural stability of the bacterial community, we examined causal relationships between structural instability (i.e., trace of interaction matrix) versus summary statistics for network topology (e.g., mean and S.D. of interaction strength), ecological properties characterizing bacterial diversity (e.g., Shannon diversity) and physicochemical environments (e.g., nutrients and salinity). To examine relationships, we applied linear analysis (temporal correlation) to determine the statistical association and nonlinear analysis (CCM) to identify causality. For linear correlation analysis, we calculated the Pearson correlation coefficient between pairs of time series and tested the significance using a stationary bootstrap that accommodates autocorrelations in time series. For nonlinear causality test, we performed CCM analysis (Chang *et al.* 2017) (See the section *Identifying causal variables by CCM* in SI Text I-1.).

*Computation*

All analyses were done with R (ver. 3.1.2). Simplex projection and CCM analyses were implemented using the rEDM (Version 1.2.3) package (Ye *et al.* 2013). The elastic-net regularization used in MDR S-map is solved by glmnet package (version 3.0). Network topological properties were computed using the igraph package (Csardi & Nepusz 2006). The computation codes of MDR S-map, as well as other analytical procedures, will be available on GitHub, upon acceptance of the manuscript.

**Results and Discussion**

*One-step forward predictability of network nodes and optimal regularization algorithms*

Prior to analyzing the reconstructed networks, we evaluated the forecasting ability of MDR S-map for the dynamics of network nodes (e.g., one-step forward forecast of future node states), as forecasting skills represent a proxy of reliability for reconstructed dynamical systems (Cenci *et al.* 2020). Compared to other methods, the MDR S-map had greater forecasting skills than other EDM methods (Table 1). In addition, for model time series, the MDR S-map outperformed all other EDM methods (Table 1), irrespective of regularization algorithms (i.e., classical (Hastie *et al.* 2009) and adaptive (Zou & Zhang 2009) elastic-net regularization). Similarly, the MDR S-map outperformed other EDM methods in empirical bacteria time series (Table 1). Specifically, the MDR S-map results, based on classical elastic-net regularization, had the best performance in both in-sample and out-of-sample forecasts. The MDR S-map results, based on adaptive elastic-net regularization, also outperformed all existing EDM methods for out-of-sample forecast, but had similar performance as the regularized S-map for in-sample forecast (Table 1). It is noteworthy that existing EDM methods (e.g. multivariate S-map) have been demonstrated to outperform other linear time series analyses on forecasting nonlinear dynamical systems (Sugihara *et al.* 1990; Deyle *et al.* 2013). Nevertheless, our proposed MDR S-map further improved forecast skills, as full information of entire networks was incorporated, whereas only partial information of sub-networks can be incorporated by the other EDM approaches that are able to accommodate only a limited number of nodes.

Although MDR S-map can effectively forecast the dynamics of network nodes, irrespective of regularization algorithms, adaptive elastic-net regularization obtained less false-positive findings in estimating interaction strengths for simulated data and thus was applied to reconstruct interaction networks throughout our analyses in this study. Comparing to the MDR S-map analysis based on adaptive elastic-net, the analysis based on classical elastic-net had slightly better forecast skills, but many more false-positive findings in estimations of interactions. False-positive rates from analyzing interactions of model networks were 1.5 and 23.3% for adaptive and classical elastic-net algorithms, respectively. The adaptive elastic-net includes additional penalty to eliminate small non-zero estimates that are potentially false positive (Zou & Zhang 2009), whereas classical elastic-net does not eliminate those nodes that have no direct interaction with the target node, but are still informative to forecast its future state (possibly through indirect interactions) (Ye *et al.* 2015). Based on this result, the best regularization algorithm optimizing one-step forecast was slightly different from that optimizing network reconstructions. As our objective is to estimate interaction strengths, we only present the results of network reconstruction based on adaptive elastic-net regularization in this work.

*Evaluating the quality of reconstructed interaction strength using simulated datasets*

MDR S-map correctly quantified the strengths of time-varying interactions (i.e., Jacobians $b_{ij}(t)$ quantifying the influence of node $j$ on $i$ at time $t$) in high-dimensional ($m = 117$) interaction networks. The estimated interaction strengths were highly consistent with theoretical expectations derived from models (Fig. 2). Such strong consistency held throughout all analyzed time points (Pearson $r$ with mean ± SD=0.79 ± 0.16), as well as for out-of-sample data (Fig. S1) and for a long-term median (Pearson $r$=0.930; Fig. S2A) obtained from temporal medians of all interaction strengths (i.e., long-term $b_{ij}$ = median($b_{ij}(t)$); $t$=1, 2, …, $n$). In contrast, the theoretical long-term medians had a weak negative relationship with interaction strengths inferred from the correlation coefficients between each pair of time series (Pearson $r$=−0.081 and $p$<0.01 in Fig. S2B), suggesting that correlation between time series provided no clear information for knowing true interaction strength in nonlinear systems, as reported (Sugihara *et al.* 2012; Freilich *et al.* 2018).

Moreover, estimations based on relative abundance (or percentage) data, a common format for biological datasets (e.g., metabarcording data (Martin-Platero *et al.* 2018)), remained effective for inferring the interaction networks with subtle biases (Fig. S3). Importantly, these estimates were also robust to random noises, including observation noises, process noises and stochastic environmental forcing (Figs. S4-S7; SI Text, *IV. Testing robustness of the MDR S-map against data noise*). In addition, the consistency between MDR S-map estimations and theoretical expectations persisted, even when some critical network nodes (e.g., dominant species ranked in top 10% of abundance) were artificially removed from the analysis (Fig. S8). Therefore, we inferred that the reconstruction of network subgraphs was still reliable, even if some critical nodes (Fig. S8) or external environmental forcing (Figs. S4-S5) were unobservable or excluded from analyses for practical reasons.

Alternate EDM approaches were unable to recover entire networks, as these methods cannot accommodate a large number of interactions in SSR models. A recent study (Ushio 2020) quantified interaction by increasing embedding dimension in regularized S-map (Cenci *et al.* 2019) with the number of causal nodes (i.e., not operating at the optimal embedding dimension). This method, although quantified individual interactions with moderate accuracy, is generally difficult to estimate network topology and stability measures (Figs. S9-S11; SI Text, *V. Importance of embedding dimension in the estimation of network topology*). Nevertheless, we recognized that our method had some limitations. For example, reconstructed networks did not reveal all existing interactions (i.e., false-negatives). That is, the number of interactions (i.e., edge number) detected in our reconstructed network may be less than that in the model network (median edge number=2098 and 889 for model and reconstructed networks, respectively) due to the difficulty in estimating weak interactions (Fig. 2).

*Evaluating the reconstructed network properties using simulated datasets*

Topological properties of network nodes

The reconstructed interaction networks preserved key topological properties of network nodes, such as strength (i.e., weighted degree), distribution of strength, and centrality, as demonstrated using an example showing the reconstructed network observed at time=950 (Fig. 3A-D). Topological indices derived at all time points also agreed with theoretical expectations (Table S1), although indices were slightly underestimated (slope <1) due to the bias of regularization (Hastie *et al.* 2009). The estimated in-strengths of each node (i.e., strength of all inward interactions to a node, Fig. 3A) had a strong positive relationship with theoretical expectations, albeit with underestimated strengths (slope <1 in Table S1) due to the bias caused by regularization (see details in SI Text, *II. Statistical properties of MDR S-map*). Similarly, the estimated out- and all-strengths (in-strength + out-strength) of each node also show strong positive associations with the theoretically expectations (Table S1). Consequently, the reconstructed cumulative distributions of in-strengths and all-strengths preserved the shape (the second or higher order moments) of theoretical distributions (all *p* >.05 for Kolmogorov-Smirnov test; Fig. 3B and Table S1). However, the shape of reconstructed out-strength distribution was not preserved, due to fewer nodes with weak out-strength (Table S1). In addition to node strength, more complex centrality indices that evaluate the importance of nodes in networks (Newman 2004), including eigenvector centrality (Fig. 3C), authority (Fig. 3D), and hub (Table S1), were also highly consistent with theoretical expectations.

Topological properties of entire network

The reconstructed interaction networks well preserved the topological properties of entire interaction networks. The topological property summarizing entire networks, such as transitivity, was well preserved (reconstructed:$0.15 \pm .01$ and true:$0.16 \pm .08$; permutation test $p$-value=0.511). Other topological properties, such as mean and standard deviation of interaction strengths, although slightly underestimated, had temporal dynamics that were highly associated with the dynamics of theoretically expected values (Fig. 3E, F). That is, these reconstructed topological properties were relatively correct; in a practical sense, due to this high consistency, we inferred that our approach was capable of monitoring topological changes of interaction networks in time.

Network stability measures

Stability measures derived from reconstructed interaction matrices (Jacobian matrices), including local Lyapunov instability (i.e., the norm of complex dominant eigenvalue; Fig. 3G) and structural instability (matrix trace; Fig. 3H), also had strong positive associations with theoretical expectations. More specifically, the estimated and theoretical local Lyapunov instability exhibited clear in-phase dynamics with concordant peaks and valleys; however, this concordance only implied a ranked but not quantitative relationship, as revealed by high Spearman and low Pearson correlations, respectively (Fig. 3G). In contrast, both ranked and quantitative relationship were preserved for structural instability (Fig. 3H). Therefore, our inferences on network stability were still reliable, although our method cannot reliably quantify all weak interactions that have long been suggested important for stability (McCann *et al.* 1998). This may be because many weak interactions are still correctly estimated (Fig. 1), which probably suffices for inferring the stability of the whole networks. Because evaluating dynamical stability of large biological systems remains challenging (Kéfi *et al.* 2019), to this end, our method warrants further investigations and offers a new research direction.

*Applications to a real-world example of bacteria community*

Keystone bacterial species identified by node centrality

As an empirical example, we reconstructed bacterial interaction networks (Movie S1) in a natural coastal environment. On average, the key OTUs exerting strong effects on others (out-strength) mainly belonged to order Flavobacteriales and order Rhodobacterales (Fig. S12) in which many members are copiotrophic species (Fuhrman *et al.* 2015) that grow rapidly in suitable environments and respond instantaneously to environmental changes. This result confirmed previous findings that Flavobacteriales are capable of degrading various polymers into more labile forms (González *et al.* 2008), which might benefit other co-existing bacterial species. Similarly, the results based on hub centrality index also revealed the importance of Flavobacteriales and Rhodobacterales as well as the other OTUs (e.g., Sphingobacteriales and Actinomycetales in Table S2), occupied the central position of interaction networks. Although the functions of these key OTUs have not been fully understood, our approach provided a promising way to identify the most critical components from a perspective of interaction network and thus guide future studies examining functions of key network nodes. It is noteworthy that the amplicon sequencing method (Martin-Platero *et al.* 2018) used to collect this bacteria dataset may have missed some important OTUs, due to incomplete DNA extraction and PCR biases. Nonetheless, the reconstructed sub-network without including every OTU may be still reliable, according to our assessment on simulated datasets (Fig. S8).

Deciphering causal mechanisms governing dynamical stability of communities

Reconstruction of high-dimensional interaction networks revealed causal mechanisms underlying dynamical stability of natural bacterial communities. Firstly, structural instability derived from the bacterial interaction networks was affected by mean interaction strength (*p*-value of CCM defined in Method, $p_{CCM}=0.027$ in Fig. 4A). Moreover, there was a marginally significant negative association between structural instability and mean interaction strength (Pearson $r=-0.250$, $p_{bootstrap}=0.098$). Because mean interaction strengths were mostly positive, this negative association implied that bacteria communities became more stable if more facilitative interactions occurred in the communities. Moreover, facilitative interactions dominated the interaction networks under productive environments, as revealed by high concentrations of chlorophyll *a* (Fig. 4B) and nutrients (e.g., silicate in Fig. 4C). Likely, nutrients facilitated the growth of primary producers producing abundant organic matters, which bring common benefits for various bacteria involved in various stages of organic decomposition and enhance the facilitative interactions. In addition, the bacterial community was more structurally stable when the Shannon diversity of bacterial community was higher (Fig. 4D), confirming previous findings of positive biodiversity effects on ecosystem stability (Chang *et al.* 2020). Apart from biological factors, structural stability was weakened by physicochemical disturbances caused by terrestrial freshwater input (revealed by reduced salinity in Fig. 4E), a common local-scale disturbance in a coastal environment (Craft 2007). These findings demonstrated a clear example elucidating key processes in natural systems through uncovering network topology and stability measures, which could not be achieved previously due to lacking a method to reliably reconstruct large dynamical networks.

*Concluding remarks*

The MDR S-map approach proposed in this study overcame the curse of dimensionality in network reconstruction. To our best knowledge, this is the first study demonstrating feasibility of quantifying interaction networks using time series data alone in high-dimensional nonlinear dynamical systems. Although the number of interactions cannot be exactly recovered, due to a lack of statistical power to differentiate some weak interactions from the absence of interaction, our method correctly quantified most critical interactions for evaluating network topology and stability. This analytical framework was applied in bacteria communities (Fig. 4) and can be easily extended to other real-world systems for searching key nodes or interactions from large networks, as long as time series of network nodes are available. Therefore, we appeal to collect high-quality time series data from various systems. As such, reconstruction of diverse types of interaction networks is expected to improve our understanding regarding complex interactions and emergent properties of large dynamical networks involving enormous numbers of interacting components.

**Acknowledgments:** We are grateful for comments from Po-Ju Ke, Cheng-han Tsai and Kazuhiro Takemoto. This study was supported by National Taiwan University, National Center for Theoretical Sciences, Foundation for the Advancement of Outstanding Scholarship, and Ministry of Science and Technology, Taiwan (to CHH).

Distinguishing error from chaos in ecological time series. *Philos. Trans. R. Soc. Lond. B.*, 330, 235-251.

Sugihara, G., May, R., Ye, H., Hsieh, C.-h., Deyle, E., Fogarty, M. *et al.* (2012). Detecting causality in complex ecosystems. *Science*, 338, 496-500.

Ushio, M. (2020). Interaction capacity underpins community diversity. *bioRxiv*, 2020.2004.2008.032524.

Ushio, M., Hsieh, C., Masuda, R., Deyle, E., Ye, H., Chang, C. *et al.* (2018). Fluctuating interaction network and time-varying stability of anatural fish community. *Nature*, 554, 360-363.

Xiao, Y., Angulo, M.T., Friedman, J., Waldor, M.K., Weiss, S.T. & Liu, Y.-Y. (2017). Mapping the ecological networks of microbial communities. *Nat. Commun.*, 8, 2042.

Ye, H., Adam Clark, Ethan Deyle, Steve Munch, Oliver Keyes, Jun Cai *et al.* (2013). rEDM: Applications of empirical dynamic modeling (EDM) from time series.

Ye, H., Deyle, E.R., Gilarranz, L.J. & Sugihara, G. (2015). Distinguishing time-delayed causal interactions using convergent cross mapping. *Sci. Rep.*, 5, 14750.

Ye, H. & Sugihara, G. (2016). Information leverage in interconnected ecosystems: Overcoming the curse of dimensionality. *Science*, 353, 922-925.

Yu, Z., Gan, Z., Huang, H., Zhu, Y. & Meng, F. (2020). The varying bacterial interactions revealed by regularized S-map. *Appl. Environ. Microbiol.*, AEM.01615-01620.

Zinzalla, G. & Thurston, D.E. (2009). Targeting protein–protein interactions for therapeutic intervention: a challenge for the future. *Future Med. Chem.*, 1, 65-93.

Zou, H. & Zhang, H.H. (2009). On the adaptive elastic-net with a diverging number of parameters. *Ann. Stat.*, 37, 1733-1751.
20

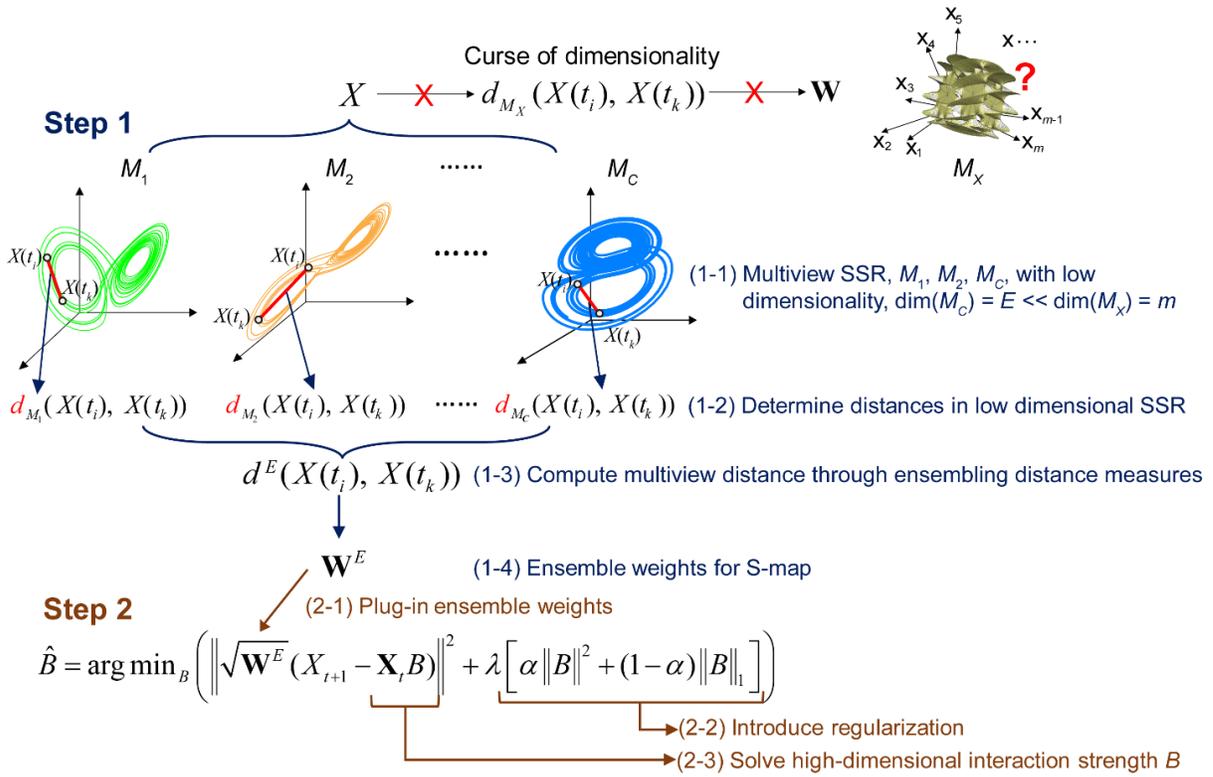

**Fig. 1. Schematic illustration for MDR S-map analysis procedure.** The MDR S-map consists of two steps. The aim of the first step is to obtain the multiview distance that operates at the optimal embedding dimension ($E$). (1-1) Neighborhood relationships [inferred from the distance, $d_M$ (.)] among high-dimensional data points $X(t)$ were recovered in numerous low-dimensional state space reconstruction (SSR; $M_1, M_2…, M_C$) (i.e., multiview SSR). Among these SSRs, (1-2) we computed distances ($L^2$ norm) among data points under the optimal embedding dimension $E$. Collecting all these distances, (1-3) we obtained the multiview distances ($d^E$) and (1-4) determined the data weights ($W^E$) for the following S-map analysis. In the second step, we estimated the high-dimensional interaction strength ($B$) by S-map, based on locally weighted least square optimization ($\mathrm{argmin}_B(\| . \|_2)^2$) with regularization ($\lambda$ [.]) incorporated. Specifically, (2-1) the weights, $w^E=\exp(-\theta d^E/\mathrm{mean}(d^E))$, derived in the first step were plugged-in the S-map optimization algorithm. (2-2) Under the constraint of regularization ($\lambda$ and $\alpha$ are the penalty factors of regularization), (2-3) we solved the high-dimensional local linear coefficients to approximate interaction strengths at each time step. Detailed definitions of each variable are reported in Methods and SI Text, *I. MDR S-map algorithm and implementation*.

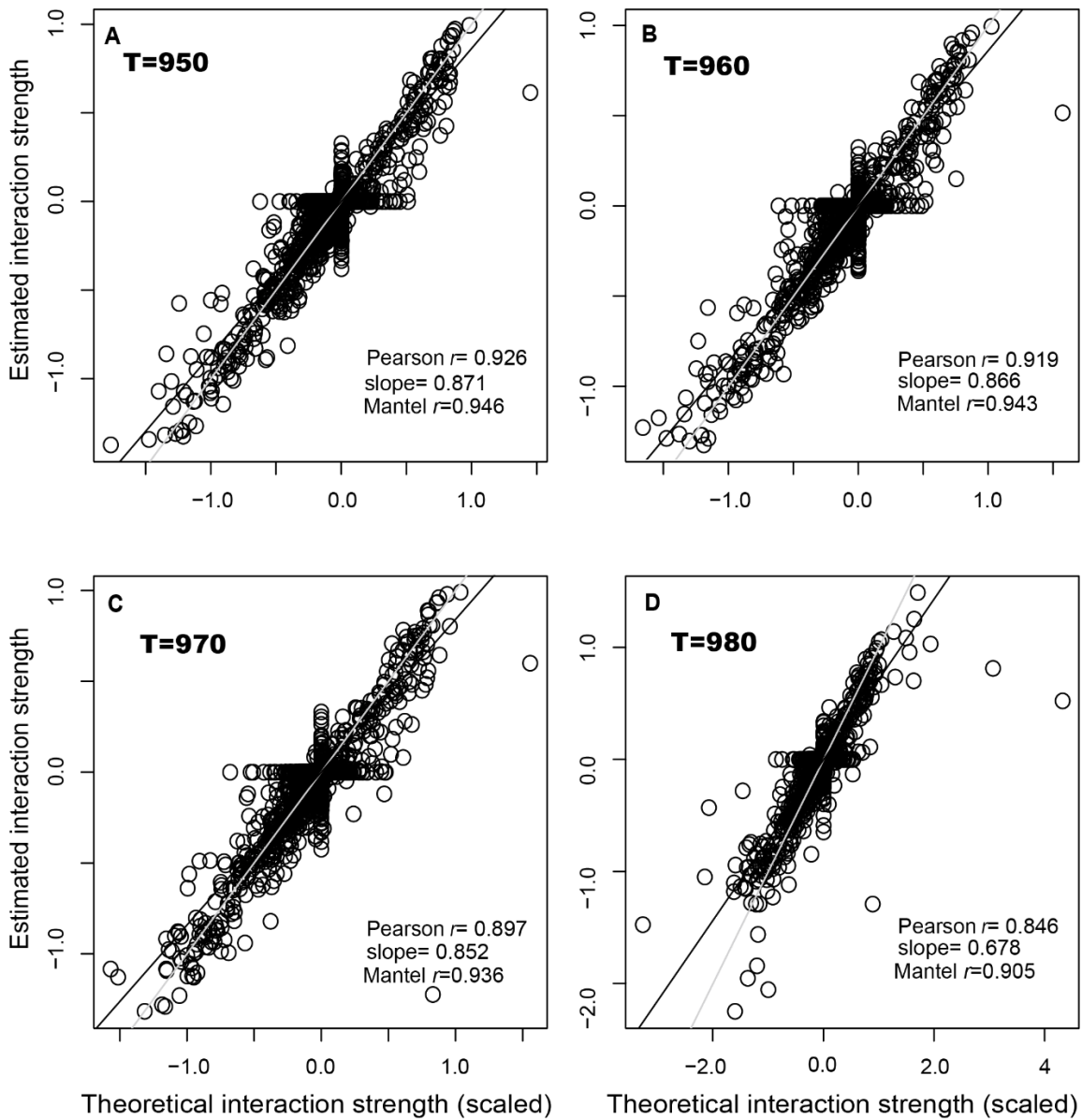

**Fig. 2. Comparisons of interaction strengths estimated by an MDR S-map, with true strengths derived from model networks.** (*A-D*) Interaction strengths of time-varying interaction networks were effectively reconstructed at every time step with high prediction skills (Pearson *r* and Mantel *r* that examine prediction skills for each corresponding pair of Jacobian elements and the entire interaction matrix, respectively); the panels are examples from four arbitrary time points at 950, 960, 970, and 980, whereas the conclusions remained for all other time points. Here, the grey lines represent the 1:1 lines, and black lines represents the slope of quantile regression based on median. The strengths were

slightly underestimated (slope<1) because the penalized algorithm used in regularization leads to minor biased (underestimated) estimators. Because strength of weak interactions cannot be precisely estimated, it caused false positive and false negative findings presented as the vertical and horizontal parts, respectively, of the crosses, near the origins.

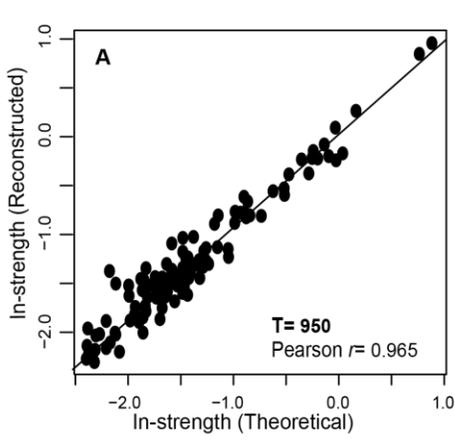
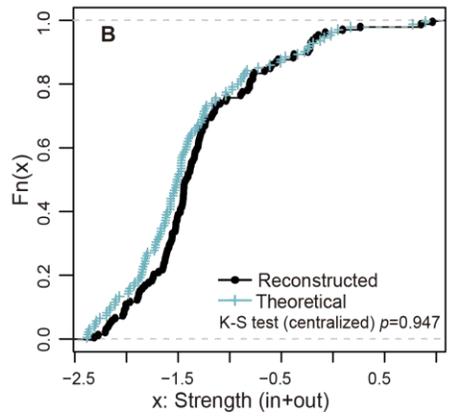
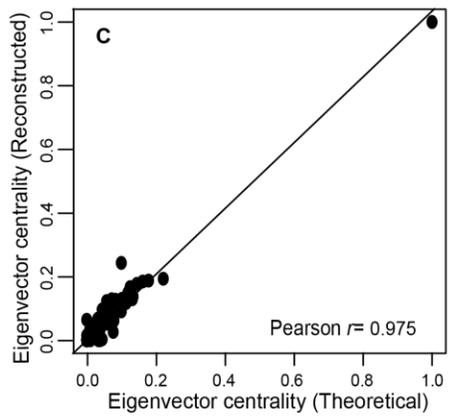
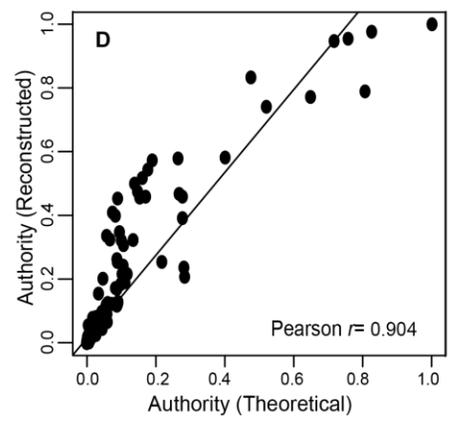
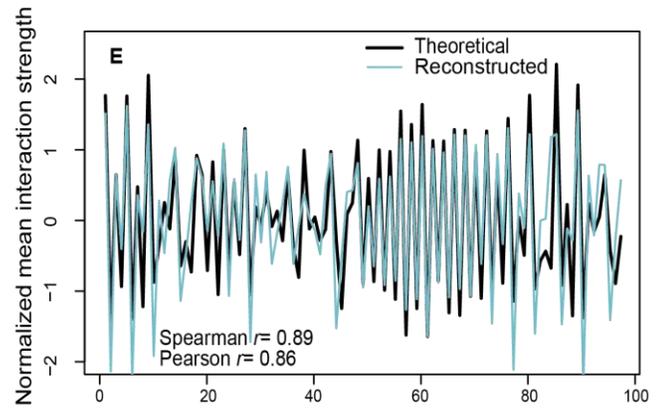
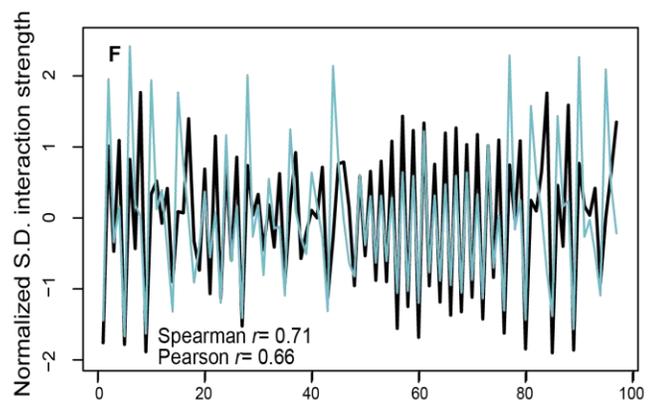
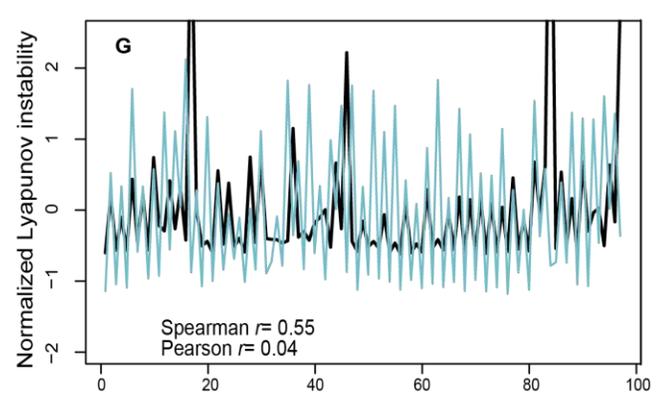
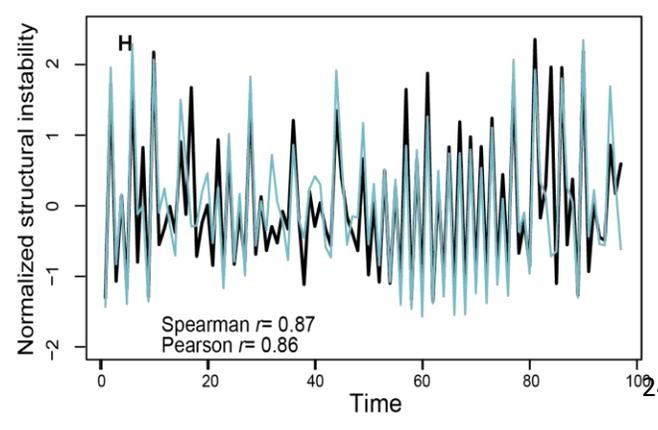

**Fig. 3**. **Preserved topological properties in the weighted, directed network reconstructed by MDR S-map.** (*A-D*) The topological properties of network nodes in the reconstructed network observed at time=950 are shown as examples. The estimated (*A*) in-strength (weighted in-degree) of each node was positively correlated with theoretical expectations. Therefore, after centralizing by mean strength, (*B*) the shape of cumulative distribution of estimated strengths was not different from that of theoretical strengths (Kolmogorov-Smirnov test *p*=0.947). Topological indices inferring the centrality of network nodes, including (*C*) eigenvector centrality and (*D*) authority, demonstrated significant positive correlations between the estimations and theoretical expectations. Results of topological properties of network nodes observed at the other time points were summarized (Table S1). For each time point, the topological properties characterizing the whole network were also quantified, including (*E*) mean and (*F*) standard deviation of interaction strength. Temporal dynamics of reconstructed topological properties had strong positive correlations with that of theoretical expectations. Similarly, indices inferring network stability, including local (*G*) Lyapunov instability and (*H*) structural instability, also demonstrated concordant temporal fluctuations in reconstructed versus theoretical networks.

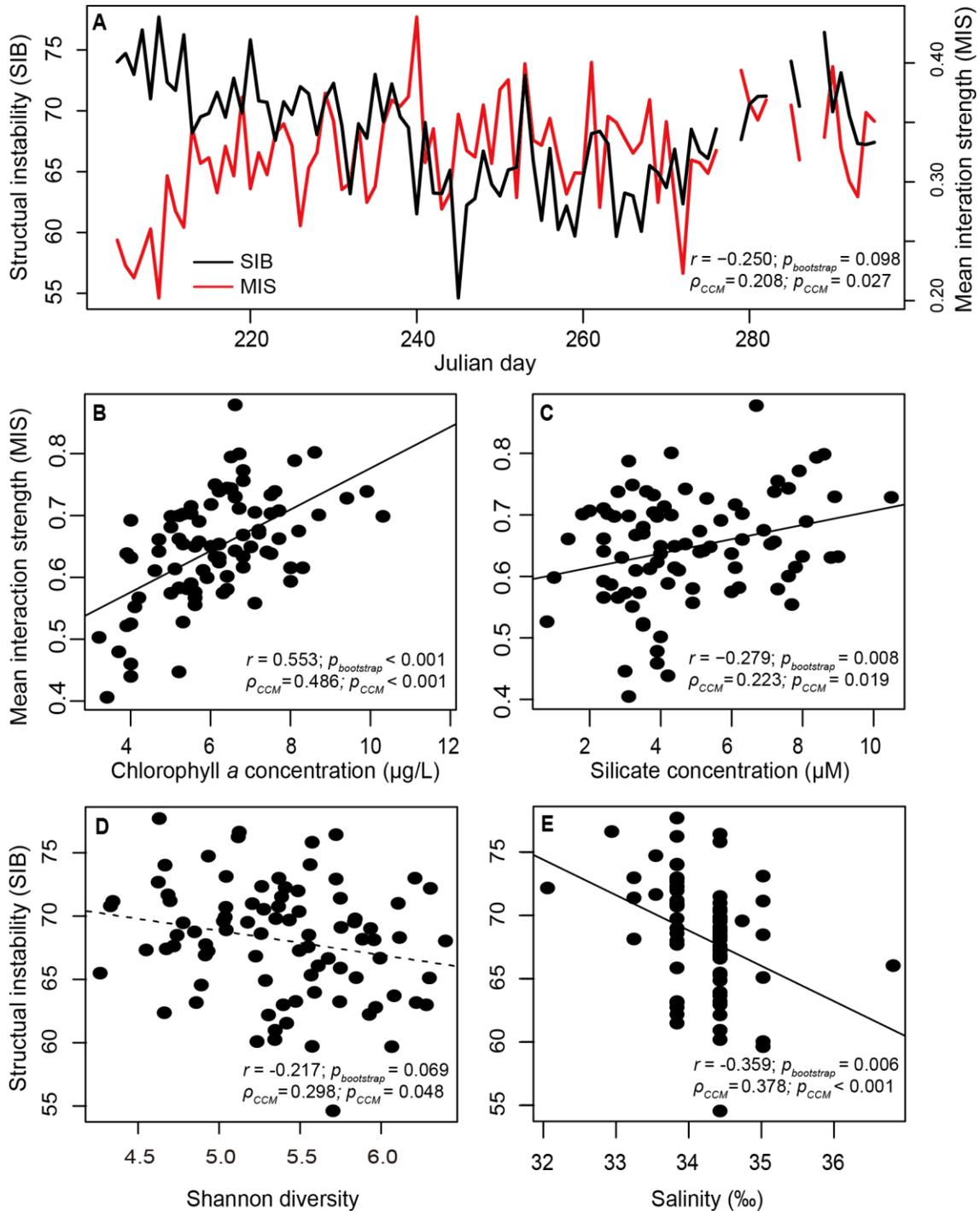

**Fig. 4. Reconstructed interaction networks from empirical daily time series revealed causal mechanisms determining structural stability in a coastal bacterial community** (*A*) There was a significant negative association between structural instability and mean interaction strength. In this system, there were more positive (facilitative) interactions

when (*B*) primary production (approximated by chlorophyll a concentration) and (*C*) nutrient concentration (e.g., silicate) were high. Moreover, structural instability decreased with increasing Shannon diversity index (*D*). In addition, local environmental disturbances caused by terrestrial freshwater input destabilized the dynamics of marine bacterial community (*E*). Correlation analysis with stationary bootstrap (*r* and $p_{bootstrap}$) and causality analysis with Convergent Cross Mapping ($\rho_{CCM}$ and $p_{CCM}$) were used to decipher the mechanisms (see Methods).

**Table 1. Comparison of forecast performance among various EDM methods.** Three types of forecast skill, Pearson *r* correlation coefficient, rMSE (root mean square error), and MAE (mean absolute error), were calculated and averaged for all individual nodes (i.e., 117 species and 135 OTUs in the simulated and empirical datasets, respectively). Here, we summarized the computed forecast skills as mean±SD. Irrespective of forecast skill types, the MDR S-map had the best performance in both leave-one-out cross-validation (in-sample) and out-of-sample forecasts. The Pearson *r* for out-of-sample forecast was not calculated, as only two test samples were used in this analysis.

| Data | Forecast skill | Type of evaluation | Univariate S-map | Multivariate S-map | Regularized S-map (classical elastic-net) | MDR S-map (classical elastic-net) | MDR S-map (adaptive elastic-net) |
|---|---|---|---|---|---|---|---|
| Model | Pearson *r* | In-sample | 0.75 ± 0.14 | 0.81 ± 0.12 | 0.83 ± 0.11 | **0.996 ± 0.01** | 0.99 ± 0.01 |
| | | Out-of-sample | - | - | - | - | - |
| | rMSE | In-sample | 0.61 ± 0.17 | 0.54 ± 0.19 | 0.52 ± 0.18 | **0.08 ± 0.05** | 0.13 ± 0.05 |
| | | Out-of-sample | 0.71 ± 0.37 | 0.67 ± 0.38 | 0.66 ± 0.39 | **0.09 ± 0.08** | 0.14 ± 0.09 |
| | MAE | In-sample | 0.47 ± 0.14 | 0.42 ± 0.14 | 0.39 ± 0.13 | **0.06 ± 0.03** | 0.09 ± 0.04 |
| | | Out-of-sample | 0.63 ± 0.35 | 0.60 ± 0.33 | 0.59 ± 0.34 | **0.08 ± 0.07** | 0.13 ± 0.08 |
| Bacterial community | Pearson *r* | In-sample | 0.48 ± 0.21 | 0.58 ± 0.18 | 0.62 ± 0.19 | **0.70 ± 0.13** | 0.63 ± 0.18 |
| | | Out-of-sample | - | - | - | - | - |
| | rMSE | In-sample | 0.90 ± 0.43 | 0.80 ± 0.16 | 0.75 ± 0.15 | **0.70 ± 0.15** | 0.78 ± 0.20 |
| | | Out-of-sample | 0.48 ± 0.58 | 0.49 ± 0.43 | 0.51 ± 0.45 | **0.41 ± 0.42** | 0.41 ± 0.43 |
| | MAE | In-sample | 0.54 ± 0.16 | 0.50 ± 0.13 | 0.47 ± 0.12 | **0.44 ± 0.11** | 0.48 ± 0.13 |
| | | Out-of-sample | 0.44 ± 0.55 | 0.45 ± 0.41 | 0.47 ± 0.42 | **0.38 ± 0.40** | 0.38 ± 0.41 |

**Supporting Information for**

5  Reconstructing large networks with time-varying interactions

Chun-Wei Chang, Takeshi Miki, Masayuki Ushio, Hsiao-Pei Lu, Fuh-Kwo Shiah, Chih-hao Hsieh*.

10  *Correspondence to: Chih-hao Hsieh; Email:chsieh@ntu.edu.tw

**This file includes:**

Supplementary Information Text I to V

I. MDR S-map algorithm and implementation

15  *I-1. Step 1: Obtaining multiview distance and weights*

*I-2. Step 2: Estimation of high-dimensional interaction strength using regularized S-map*

*I-3. Parameter selection based on cross-validation*

*I-4. State dependency of estimated interaction strength*

20  *I-5. Dimensional difference in the two-step procedure of MDR S-map*

II. Statistical properties of MDR S-map

*II-1. Statistical properties of interaction strengths estimated by ordinary and regularized S-map*

*II-2. Statistical properties of MDR S-map based on percentage data*

25  III. Parameterization of theoretical model

IV. Testing robustness of the MDR S-map against data noise

*IV-1. Random noises generation*

*IV-2. Robustness of MDR S-map to noises*

V. Importance of embedding dimension in the estimation of network topology

Figures S1 to S12
Tables S1 to S2
Legends for Movies S1
SI References

**Other supplementary materials for this manuscript include the following:**
Movies S1

# Supplementary Information Text

## I. MDR S-map algorithm and implementation

This section offered the detail algorithm for the implementation of multiview distance regularized S-map (MDR S-map), which can be achieved in two steps.

*I-1. Step 1: Obtaining multiview distance and weights*

The implementation of multiview SSR

In the first step, we determined the multiview distance that depicted the neighbors of a high-dimensional system state via the ensemble of numerous distances measured in various low-dimensional state space reconstructions at the optimal embedding dimension ($E$). Practically, the optimal embedding dimension ($E_i$) for a network node, $x^{(i)}$ was obtained using univariate simplex projection (Sugihara & May 1990) that optimized a one-step forward forecast, predicting the future states of $x^{(i)}$. With the determined optimal embedding dimension, numerous low-dimensional SSRs were built by embedding various combinations of causal variables and their time-lags (maximal lag=3, following multiview SSR (Ye & Sugihara 2016)). For instance, to make one-step forward forecast on one of the $m$ network nodes $x^{(i)}$, we considered a subset of variables $\{x^{(i_1)}, x^{(i_2)}, ..., x^{(i_{E_i})}\}$ from all the $q_i$ causal variables of $x^{(i)}$. Generally, the number of causal variables affecting $x^{(i)}$ ($q_i$) and network dimensionality ($m$) were much larger than the optimal embedding dimension ($E_i$) in large dynamical systems, i.e., $m \geq q_i >> E_i$.

Identifying causal variables by CCM

Practically, all causal variables from $m$ network nodes are determined using convergent cross-mapping (CCM) (Sugihara *et al.* 2012). The significance of CCM causality was obtained by testing the convergence of crossing-mapping from the node $x^{(i)}$ to all other $m$-1 network nodes (Chang *et al.* 2017). The convergence, if any, can be recognized as the improvement of cross-mapping skills $\rho(L)$ (correlation coefficients between observations and predictions) when increasing time series library, $L$. Specially, we tested convergence by testing the significance of: 1) monotonic increasing trend in $\rho(L)$ using Kendall's $\tau$ test; and 2) converged cross-mapping skill $\rho(L_{max})$ using a Student's $t$-test, where the significance of CCM requires that both tests are significant (i.e., $\max(p_{\text{Kendall-test}}, p_{t\text{-test}}) < \alpha = 0.05$) (Chang *et al.* 2017).

Multiview distance and weights derived based on the ensemble approach

With the determined causal variables and the optimal embedding dimension of the target node $i$, we selected various subsets of causal variables to build numerous low-dimensional SSR ($M_c$; $c$ denotes any one of the combinations consisting of the causal variables) operated at the optimal embedding dimension, $E_i$. For all these SSR reconstructed for node $i$, we calculated the distance ($L^2$ norm) between every pair of embedded states observed at various time points, e.g., $X^c(t_u) = (x^{(c,i_1)}, x^{(c,i_2)}, ..., x^{(c,i_{E_i})})(t_u)$ and $X^c(t_v) = (x^{(c,i_1)}, x^{(c,i_2)}, ..., x^{(c,i_{E_i})})(t_v)$, where $x^{(c,i_k)}$ the $k$-th causal variable selected for the node $i$ in the variable combination $c$, respectively. $\forall t_u, t_v \in T: \{t_1, t_2, ... t_N\}$ and $t_u \neq t_v$, using their Euclidean distance $d(X^c(t_u), X^c(t_v))$. Finally, we collected the distances measured under various SSR, $M_c$ and then calculated the ensemble multiview distance, $d^E(X_m(t_u), X_m(t_v))$, as the weighted average among all these distances, i.e., $d^E(X_m(t_u), X_m(t_v)) = \sum_{\forall c} w_c d(X^c(t_u), X^c(t_v))$, to approximate the distance measured among high-dimensional system states, $X_m(t) = (x^{(1)}, x^{(2)}, ..., x^{(m)})(t)$. Here, the weight, $w_c$, is proportional to the forecast skill that evaluates the performance of one-step forward forecasts for the future values of $x^{(i)}$ under $M_c$, using the correlation coefficients ($\rho$) between model forecast and observations, (i.e., $w_c \propto \rho(M_c), \sum_{\forall c} w_c = 1$). In practical implementation, we randomly generated 10000 SSR from combinations of causal variables and retained only the top 100 SSR with the highest one-step forecast skills ($c$=1, 2, …, 100). This strategy considers the computational efficiency, since the number of low-dimensional SSR, $M_c$, also grows combinatorically with the number of causal variables $q$. Finally, we prepared a $N \times N$ multiview distance matrix that collected all pairwise ensemble distances for the analyses in the next step. Again, this distance matrix was determined at the optimal embedding dimension for each node.

Remarks of Step 1

The computation of multiview distance under numerous low-dimensional SSR, i.e., $d^E(X_m(t_u), X_m(t_v))$, recovered the information of high-dimensional SSR, but avoided the computation of distance measures using all network nodes, i.e., $d(X_m(t_u), X_m(t_v))$. This step is critical, as mean distance among data points increases with data dimensionality, or equivalently, the sample size required to maintain constant mean distance among randomly generated data points grows exponentially with increasing dimensionality, known as the curse of dimensionality (Bellman 1957; Hall *et al.* 2005). As a consequence, the distance measured under the high-dimensional state space may not reliably preserve correct neighboring relationships among high-dimensional states, i.e., $X_m(t)$ that are required for SSR-based approaches (Sugihara *et al.* 1990; Kennel *et al.* 1992; Shalizi 2006).

*I-2. Step 2: Estimation of high-dimensional interaction strength using regularized S-map*

Quantitative definition of interaction strength in S-map

The second step is to estimate interaction strengths based on the neighboring relationships determined in the first step (i.e., multiview distance and weights). In this step, we adopted an intuitive definition of interaction strength: the strength of node $x^{(j)}$ affecting node $x^{(i)}$ at time $t$ is the response of $x^{(i)}$ at $t$+1 to subtle changes of $x^{(j)}$ at previous time step $t$, i.e., $\partial x^{(i)}(t+1)/\partial x^{(j)}(t)$. In general, we could always identify a time-recursive function, $F_i$ that linked the response of $x^{(i)}(t+1)$ to all interacting nodes observed at previous time step $t$,

$$x^{(i)}(t+1) = F^{(i)}(X_m(t)), \forall t \in T: \{t_1, t_2, ..., t_N\}, X_m(t) = (x^{(1)}(t), x^{(2)}(t), ..., x^{(m)}(t))^T,$$

$F^{(i)}: \mathbb{R}^m \to \mathbb{R}^1$, and $i = 1, 2, ..., m$. Then, the interaction strength of $x^{(j)}$ affecting $x^{(i)}$ at time $t$ is equivalent to the partial derivative of the time-recursive function $F^{(i)}$ with respect to $x^{(j)}$, $\partial F^{(i)}(X_m(t))/\partial x^{(j)}$, evaluated at time $t$ (i.e., the Jacobian of $F^{(i)}$). The derived interaction strength based on time-recursive function quantifies the interaction that occurred at the time scale between two adjacent time points (e.g., sampling interval of time series). Therefore, changing the sampling interval alters the time-recursive function and thus infers the interaction strength at various time scales.

S-map algorithm based on local linear approximation

In reality, the time-recursive function $F^{(i)}$ is unknown and highly complicated and thus cannot be parametrically specified. To solve this problem, S-map (Sugihara *et al.* 1990; Deyle *et al.* 2016) approximates $F^{(i)}$ locally using linear combinations of embedded variables observed at each time point $t_k$, i.e., $F^{(i)}(X_m(t_k)) \simeq f_{t_k}^{(i)}(X_m(t_k)) = X_m^T(t_k)B_{t_k}^{(i)} = \sum_{s=0}^{m} b_{is}(t_k)x^{(s)}(t_k)$, $\forall t_k \in T: \{t_1, t_2, ..., t_N\}$, where $x^{(0)} = 1$ is the intercept term and $f_{t_k}^{(i)}$ is the function that locally approximates the unknown function $F^{(i)}$ at time $t_k$. Based on this local linear approximation, the interaction strengths of $x^{(j)}$ affecting $x^{(i)}$ at time $t_k$, $\partial F^{(i)}(X(t_k))/\partial x^{(j)}(t_k)$, can be approximated by the Jacobians of these locally approximated linear functions, (i.e., local linear coefficients), $\partial f_{t_k}^{(i)}(X_m(t_k))/\partial x^{(j)}(t_k) = b_{ij}(t_k)$. Therefore, we can estimate the interaction strengths (Jacobians) by estimating these local linear coefficients, $b_{ij}(t_k)$. To do so, S-map proposed a local weighted least square estimator to solve the local linear coefficients,

$$\hat{B}_{t_k}^{(i)} = \arg\min_B \left\| \sqrt{\mathbf{W}_{t_k}^{(i)}} (Y^{(i)} - \mathbf{X} B_{t_k}^{(i)}) \right\|^2 \quad \text{--------(eq. 1)},$$

where $\mathbf{X} = (X^{(1)}, X^{(2)}, ..., X^{(m)})$ is a $N \times m$ data matrix collecting the time series of all network nodes; $X^{(i)} = (x^{(i)}(t_1), x^{(i)}(t_2), ..., x^{(i)}(t_N))^T$ and $Y^{(i)} = (x^{(i)}(t_2), x^{(i)}(t_3), ..., x^{(i)}(t_{N+1}))^T$ are $N \times 1$ vectors representing the current and one-step forward time series data of the node $x^{(i)}$, respectively. $\mathbf{W}_{t_k}^{(i)}$ is a $N \times N$ diagonal ensemble weight matrix in which each diagonal entity is the weight obtained from the exponential decay function of Euclidean distance, $d(X_m(t_q), X_m(t_k))$, measured under SSR, $w_q = \exp\left[-\theta \frac{d(X_m(t_q), X_m(t_k))}{\bar{d}}\right] \geq 0$, $\forall t_q \in T : \{t_1, t_2, ..., t_N\}$, where $\theta$ is a state-dependency (nonlinearity) parameter and $\bar{d} = \frac{1}{N} \sum_{\forall q} d(X_m(t_q), X_m(t_k))$ is the mean distance. Here, the target point is weighted by zero ($w_{kk}=0$) to exclude the target data points from the estimating equation (i.e., leave-one-out cross-validation). The Jacobians are solved for each time point; thus, a time series of interaction strengths (i.e. time-varying interaction network) are estimated.

Solving high-dimensional interaction strength using regularization

As explained in the first step, the high dimensionality of network ($m$) is vulnerable to the curse of dimensionality. To tackle this issue, we replaced the distance $d$ measured in high-dimensional space (dimension=$m$) by ensemble multiview distances $d^E$ (derived in the first step) measured at the optimal embedding dimension (dimension=$E$) to determine the local weight matrix $\mathbf{W}_{t_k}^E$. Furthermore, the optimization based on Eq. 1 suffered from overfitting problems and has no unique solution if the dimensionality ($m$) is larger than the time series length ($N$) (Bühlmann & Van De Geer 2011). Thus, we constrained the optimization by introducing regularization (e.g., elastic-net (Cenci *et al.* 2019)) to estimate the high-dimensional coefficients, $\hat{B}_{t_k}^{(i)}$,

$$\hat{B}_{t_k}^{(i)} = \arg\min_B \left( \left\| \sqrt{\mathbf{W}_{t_k}^E}(Y^{(i)} - \mathbf{X}B_{t_k}^{(i)}) \right\|^2 + \lambda \left[ \alpha \left\| B_{t_k}^{(i)} \right\|^2 + (1-\alpha) \left\| B_{t_k}^{(i)} \right\|_1 \right] \right) \text{---------(eq. 2)}.$$

Here, $\lambda$ is the penalized factor, and $\alpha$ is the adjusted parameter balancing the regularization using $L^1$ ($\|.\|_1$) or $L^2$ ($\|.\|$) norm of the parameter vector $B_{t_k}^{(i)}$. The penalization on $L^1$ norm automatically shrinks parameters $b_{ij}(t_k)$ to zero (Tibshirani 1996), if some variables $x^{(j)}$ have no or very weak effect on $x^{(i)}$ at time $t_k$. Based on the algorithm of parameter shrinkage, the embedded variables do not need to be determined *in prior* as in existing multivariate EDM methods; instead, the variables having substantial effects are automatically (estimated $b_{ij} \neq 0$) selected at each time point. To eliminate potential false positives, we applied adaptive elastic-net regularization that further shrinks the estimated Jacobians (Zou & Zhang 2009). In summary, the proposed two-step procedure of MDR S-map quantified high-dimensional interaction strengths and reconstructed interaction networks.

*I-3. Parameter selection based on cross-validation*

Practically, the solution of Eq. 2 depends on nonlinearity parameter $\theta$, penalized factor $\lambda$, and adjusted parameter $\alpha$ in the MDR S-map algorithm. The best parameter combination ($\theta_i$, $\lambda_i$, $\alpha_i$) for each node $i$ and estimated interaction strengths are those that minimize rMSE of the one-step forecast on $x^{(i)}(t_k+1)$. More specifically, the parameters were determined by leave-one-out cross-validation, in which data points collected at the testing time step $t_k$ were excluded (weighted by zero) from the dataset ($Y^{(i)}$ and $\mathbf{X}$) to obtain a one-step forecast on $x^{(i)}(t_k+1)$, $\forall t_k \in T: \{t_1, t_2, \ldots, t_N\}$. In addition, we keep self-regulation effect of each node ($\hat{\beta}_{ii}$ in Eq. 2) in the final estimation, which the estimation of $\hat{\beta}_{ii}$ underwent no penalization, i.e., $\lambda_i = 0$ but $\hat{\beta}_{ii} \neq 0$. To test the generalization ability of our methods, the selected parameters were applied for fitting those time series data involved in the cross-validation (i.e., library or in-sample) and also for evaluating data not involved in parameter selections (i.e., out-of-sample) (Cenci *et al.* 2019). Finally, we repeated all these computations for every node and eventually derived the estimated Jacobian matrix (i.e., interaction matrix) at time step *t*, in which each entity $J_{ij} = \hat{\beta}_{ij}$ was estimated as S-map coefficient representing the strength of node *j* affecting node *i*.

In addition to cross-validation, we then applied out-of-sample prediction to evaluate the performance of S-map model trained by library dataset (Cenci *et al.* 2019). Specifically, we left the last two data points as out-of-samples and excluded them from the library data set used for cross-validations and model fittings, i.e., out-of-samples were not a part of the library data set ($Y^{(i)}$ and $\mathbf{X}$ in the eq. 2.). Estimations of out-of-samples interaction strengths were achieved by plugging-in their neighborhood relationships with respective to library data (i.e., the weights derived from ensemble distance, $d^E(X_{\text{library}}, X_{\text{out-of-sample}})$) when solving Eq. 2. It is noteworthy that the out-of-samples predictions in interaction strengths highlighted the capacity of MDR S-map to predict the one-step forward future state of entire interaction networks.

*I-4. State dependency of estimated interaction strength*

The estimation of interaction strengths by MDR S-map is based on Jacobian $\partial x^{(i)}(t+1) / \partial x^{(j)}(t)$, considering the overall effects of abundance changes of nodes *j* on the abundance of nodes *i*. This measurement changed through time, i.e., time-varying, if either one of these following scenarios occurs: i) the abundance of $x^{(i)}$ or $x^{(j)}$ is time-varying; or ii) the per capita effects of $x^{(j)}$ are time-varying. Scenario (i) usually occurs in nonlinear dynamical systems, whereas Scenario (ii) is often observed in empirical systems, but often ignored in many theoretical analysis for the reason of simplicity. Certainly, a combination of (i) and (ii) are also possible. Although the per capita effects can be isolated in simple theoretical systems via the scaling of abundances (Berlow *et al.* 2004), we do not recommend the isolation of per capita effects from empirically estimated overall effects, as the scaling of abundance is numerically unstable, especially when the observed abundance is low.

*I-5. Dimensional difference in the two-step procedure of MDR S-map*

Unlike existing EDM methods, MDR S-map uses ensemble neighborhood relationships obtained from various views of low-dimensional SSRs (i.e., multiview distance and weights, $d^E$ and $\mathbf{W}^E$, obtained at the optimal dimension, $E$) but accesses high-dimensional estimators of interaction strengths using regularization. As such, we constructed high-dimensional, time-varying interaction networks in large dynamical systems. It is noteworthy that the dimensionality used in the first step ($E$) and the second step ($m$) were different. Under the optimal embedding dimension at the first step, the reconstructed manifold was the topological invariant to the original manifold (Sauer *et al.* 1991), wherein systems situated in similar contexts (i.e., neighbors) will have similar dynamics (Sugihara & May 1990). Thus, identifying neighbors on the SSR with a fixed optimal embedding dimension is a critical concern. Although there are many nodes (at most $m$) interacting with the target node ($x^{(i)}$) we are concerned about, it is very likely that the effective dimensionality is not so high, as many effects are redundant; consequently, the optimal embedding dimension $E$ is always less than $m$. Therefore, we do *not* need to limit the number of potential interacting variables in Eq. 2 in the statistical inferences of interaction strengths, as the number of interacting nodes is not equal to the dimensionality of dynamical attractor reconstructed from the view of the target node $x^{(i)}$. This reasoning, setting the dimensionality differently for different parts of algorithm, is the main difference between our method and existing EDM methods.

MDR S-map based on the ensemble approach (i.e., the first step; Fig. 1) effectively determines the neighboring relationship among high-dimensional data points at the optimal embedding dimension. The neighboring relationship is critical to investigating the dynamical behavior of nonlinear systems (Sugihara & May 1990), but cannot be directly inferred from the distance measured in high-dimensional state space, as the curse of dimensionality makes the median distance among data points increase with their dimensionality (Bellman 1957). Combining an ensemble approach with the regularization (i.e., the second step; Fig. 1) makes the interaction strengths of high-dimensional networks explicitly quantified. This quantification needs no prior knowledge or complicated procedures to select variables as in previous studies (Deyle *et al.* 2016; Suzuki *et al.* 2017; Cenci & Saavedra 2019). Instead, our algorithm based on regularization automatically determines interacting components at each time point from all network nodes and thus preserves key topological properties of large networks (Fig. 2 and 3).

## II. Statistical properties of MDR S-map

*II-1. Statistical properties of interaction strengths estimated by ordinary and regularized S-map*

In this section, we derived the statistical expectations of original and regularized S-map coefficients. Considering a dynamical system presented by time-recursive function using embedded state variables,

$$x^{(i)}(t+1) = F^{(i)}(X_m(t)), \forall t \in (1,2,...,n), F_i : \mathbb{R}^m \to \mathbb{R}^1$$

Here, $X_m(t)$ is a point at native state space (not through taking lags) observed at time $t$, $X_m(t)=(x^{(1)}(t), x^{(2)}(t), ..., x^{(m)}(t))^T$. While $F_i$ is nonlinear and thus difficult to estimate reliably; thus, we estimated the function nonparametrically (Perretti *et al.* 2013) using local linear approximation of all embedded variables at each time step $t$, such that

$$F^{(i)}(X_m(t)) \simeq f_t^{(i)}(X_m(t)) = X_m^T(t)B_t^{(i)} = \sum_{s=0}^{m} b_{is}(t)x^{(s)}(t), \forall t_k \in T : \{t_1, t_2, ..., t_N\}.$$

Statistically, the local linear coefficients, $B_t^{(i)}$, can be estimated by applying weighted-least square locally at each data point $X_{tk}$, which repeatedly solves the least square optimization using the whole dataset $\mathbf{X}_t$ (dim($\mathbf{X}_t$)=$N*m$), albeit with different weights for different data points, depending on how close the library data points $X_{tq}$ are to the target points $X_{tk}$ (inferred from $d(X_{tq}, X_{tk})$) in State Space Reconstruction (SSR). Then, the estimator of $B_{t_k}^{(i)}$ at every time step can be solved by local weighted least square,

$$\hat{B}_{t_k}^{(i)} = \arg\min_B \left\| \sqrt{\mathbf{W}_{t_k}^{(i)}}(Y^{(i)} - \mathbf{X}B_{t_k}^{(i)}) \right\|^2 = (\mathbf{X}^T\mathbf{W}_{t_k}^{(i)}\mathbf{X})^{-1}\mathbf{X}^T\mathbf{W}_{t_k}^{(i)}Y^{(i)},$$

where, $Y^{(i)} = (x^{(i)}(t_2), x^{(i)}(t_3), ..., x^{(i)}(t_{N+1}))^T$, $\mathbf{W}_{t_k} = \text{diag}(w_1, w_2, ..., w_n)$, and $w_q = \exp\left[-\theta \frac{d(X(t_q), X(t_k))}{\bar{d}}\right]$. This estimator is an unbiased estimator because

$$E(\hat{B}_{t_k}^{(i)} | \mathbf{X}) = (\mathbf{X}^T\mathbf{W}_{t_k}^{(i)}\mathbf{X})^{-1}\mathbf{X}^T\mathbf{W}_{t_k}^{(i)}E(Y^{(i)}) = (\mathbf{X}^T\mathbf{W}_{t_k}^{(i)}\mathbf{X})^{-1}\mathbf{X}^T\mathbf{W}_{t_k}^{(i)}\mathbf{X}B_{t_k}^{(i)} = B_{t_k}^{(i)}.$$

As explained in the Methods section, due to high dimensionality, the weight matrix in our method was not directly calculated from all the variables using the whole dimension. Instead, we determined the weight matrix based on the ensembles collected from low-dimensional SSR under the optimal embedding dimension (see Methods). Nevertheless, the replacement of multiview weight matrix will not make this estimator biased.

Next, we extended the statistical derivation of S-map coefficients that accommodated the regularization using elastic-net. In regularization, we included the penalty factor ($\lambda$) that added constraints on least square optimization and shrunk the estimated parameters.

$$\hat{B}_{t_k}^{(i)} = \arg\min_B \left( \left\| \sqrt{\mathbf{W}_{t_k}^{(i)}} (Y^{(i)} - \mathbf{X} B_{t_k}^{(i)}) \right\|^2 + \lambda \left[ \alpha \left\| B_{t_k}^{(i)} \right\|^2 + (1-\alpha) \left\| B_{t_k}^{(i)} \right\|_1 \right] \right)$$

Here, we only demonstrated a special case of elastic-net regularization using ridge regression (i.e., $\alpha=1$), in which the estimator had close form.

$$\hat{B}_{t_k}^{(i)} = \arg\min_B \left( \left\| \sqrt{\mathbf{W}_{t_k}^{(i)}} (Y^{(i)} - \mathbf{X} B_{t_k}^{(i)}) \right\|^2 + \lambda \left\| B_{t_k}^{(i)} \right\|^2 \right) = (\mathbf{X}^T \mathbf{W}_{t_k}^{(i)} \mathbf{X} + \lambda \mathbf{I}_n)^{-1} \mathbf{X}^T \mathbf{W}_{t_k}^{(i)} Y^{(i)}$$

Obviously, this estimator was biased and this biased estimator is always underestimated, as the penalty factor $\lambda$ is positive.

$$\mathrm{E}(\hat{B}_{t_k}^{(i)} | \mathbf{X}) = (\mathbf{X}^T \mathbf{W}_{t_k}^{(i)} \mathbf{X} + \lambda \mathbf{I}_n)^{-1} \mathbf{X}^T \mathbf{W}_{t_k}^{(i)} \mathbf{X} B_{t_k}^{(i)} = B_{t_k}^{(i)} - \lambda (\mathbf{X}^T \mathbf{W}_{t_k}^{(i)} \mathbf{X} + \lambda \mathbf{I}_n)^{-1} B_{t_k}^{(i)}$$

Although this regularized estimator is biased (i.e., slightly lower accuracy), it provides greater precision (lower variance) by avoiding overfitting, making the fitted S-map model have better generalization ability for predicting the dataset out of library (out-of-sample). This purpose of improving forecast precision was addressed in the previous study (Cenci *et al.* 2019). Here, our purpose for using regularization was to introduce additional constraints on least square optimization that made estimations of high-dimensional interaction strengths possible. That is, if there is no regularization, the least square optimization cannot obtain unique solutions (e.g., infinite solutions) of the S-map coefficients when the number of data points ($N$) is less than the number of variables ($m$); this difficulty hinders reconstruction of high-dimensional interactional networks (Tibshirani 1996).

*II-2. Statistical properties of MDR S-map based on percentage data*

In this section, we derived the statistical properties of S-map coefficients, based on the percentilized time series data ($P_t$). Firstly, we formulated the time-recursive function $G^{(i)}$ used in the S-map, but used percentage data instead.

$$p^{(i)}(t+1) = G^{(i)}(P(t)) \simeq P^T(t)\Upsilon_t^{(i)}$$

$$p^{(i)}(t) = \frac{x^{(i)}(t)}{\sum_{\forall k} x^{(k)}(t)} = c_t^{-1} x^{(i)}(t) \text{ and } c_t = \sum_{\forall k} x^{(k)}(t)$$

Again, the time-recursive function is locally approximated by linear function, with linear coefficients $\Upsilon_t^{(i)}$. Similarly, we solved the weighted-least square optimization to obtain the estimator of local linear coefficients,

$$\hat{\Upsilon}_{t_k}^{(i)} = \arg\min_\Upsilon \left\| \sqrt{\mathbf{W}_{t_k,p}^{(i)}} (Y_p^{(i)} - \mathbf{P}\Upsilon_{t_k}^{(i)}) \right\| = \arg\min_\Upsilon \left\| \sqrt{\mathbf{W}_{t_k,p}^{(i)}} (\mathbf{C}_{t+1}^{-1} Y^{(i)} - \mathbf{C}_t^{-1}\mathbf{X}\Upsilon_{t_k}^{(i)}) \right\|$$

, where $\mathbf{C}_t = \text{diag}(c_{t_1}, c_{t_2}, ..., c_{t_N})$. Assuming the weight matrix $\mathbf{W}_{t_k,p}^{(i)} = \mathbf{D}_w \mathbf{W}_{t_k}^{(i)}$ derived from proportional data, with $\mathbf{D}_w = \text{diag}(1+\delta_1, 1+\delta_2, ..., 1+\delta_n)$ is different from the original weight matrix with some derivations $\delta_s$.

$$\hat{\Upsilon}_{t_k}^{(i)} = \arg\min_\Upsilon \left\| \sqrt{\mathbf{W}_{t_k,p}^{(i)}} (\mathbf{Q}_t Y^{(i)} - \mathbf{X}\Upsilon_{t_k}^{(i)}) \right\| = (\mathbf{X}^T \mathbf{W}_{t_k,p}^{(i)} \mathbf{X})^{-1} \mathbf{X}^T \mathbf{W}_{t_k,p}^{(i)} \mathbf{Q}_t Y^{(i)}, \text{ where } \mathbf{Q}_t \triangleq \mathbf{C}_{t+1}^{-1}\mathbf{C}_t,$$

Then, we derive the expectation of estimator, $\hat{\Upsilon}_{t_k}^{(i)}$ as

$$E(\hat{\Upsilon}_{t_k}^{(i)} | \mathbf{X}) = (\mathbf{X}^T \mathbf{W}_{t_k,p}^{(i)} \mathbf{X})^{-1} \mathbf{X}^T \mathbf{W}_{t_k,p}^{(i)} \mathbf{Q}_t E(Y^{(i)}) = (\mathbf{X}^T \mathbf{W}_{t_k,p}^{(i)} \mathbf{X})^{-1} (\mathbf{X}^T \mathbf{W}_{t_k,p}^{(i)} \mathbf{Q}_t \mathbf{X}) B_{t_k}^{(i)}$$

$$= (\mathbf{X}^T \mathbf{W}_{t_k,p}^{(i)} \mathbf{X})^{-1} (\mathbf{X}^T \mathbf{W}_{t_k,p}^{(i)} (\mathbf{I}_n + \mathbf{D}_Q) \mathbf{X}) B_{t_k}^{(i)} = B_{t_k}^{(i)} + \underbrace{(\mathbf{X}^T \mathbf{W}_{t_k,p}^{(i)} \mathbf{X})^{-1} (\mathbf{X}^T \mathbf{W}_{t_k,p}^{(i)} \mathbf{D}_Q \mathbf{X}) B_{t_k}^{(i)}}_{\text{biased due to percentilization}}$$

$\mathbf{D}_Q \triangleq \text{diag}(\frac{c_{t_1} - c_{t_1+1}}{c_{t_1+1}}, \frac{c_{t_2} - c_{t_2+1}}{c_{t_2+1}}, ..., \frac{c_{t_n} - c_{t_n+1}}{c_{t_n+1}})$, $\hat{\Upsilon}_{t_k}^{(i)}$ is unbiased only if $\mathbf{Q}_t$ is an identity matrix (or $\mathbf{D}_Q$ is a zero matrix). Based on this formulation, the estimator may be biased because there are unequal weightings operating on $X_{t+1}^{(i)}$ and **X** by $\mathbf{C}_{t+1}^{-1}$ and $\mathbf{C}_t^{-1}$, respectively. We inferred that for the estimation for percentage data to be reliable, a statistical requirement was that fluctuations in total biomass ($c_t$) between adjacent time points are not too strong. It is also noteworthy that the difference in weight matrices, $\mathbf{W}_{tk}$ and $\mathbf{W}_{tk,p}$, itself, will not bias the estimation, unless $c_{t+1}^{-1}$ and $c_t^{-1}$ are very different. Finally, we extend this derivation by including the penalty factor ($\lambda$) introduced by regularization based on $\alpha=1$.

$$\mathrm{E}(\hat{\Upsilon}_{t_k}^{(i)} \mid \mathbf{X}) = B_{t_k}^{(i)} - \underbrace{\lambda (\mathbf{X}^T \mathbf{W}_{t_k,p}^{(i)} \mathbf{X} + \lambda \mathbf{I}_n)^{-1} B_{t_k}^{(i)}}_{\text{biased due to regularization}} + \underbrace{(\mathbf{X}^T \mathbf{W}_{t_k,p}^{(i)} \mathbf{X} + \lambda \mathbf{I}_n)^{-1} (\mathbf{X}^T \mathbf{W}_{t_k,p}^{(i)} \mathbf{D}_Q \mathbf{X}) B_{t_k}^{(i)}}_{\text{biased due to percentilization+regularization}}$$

The derived estimation is still biased. The bias comes from two sources: One is caused by parameter shrinkage controlled by penalty factor ($\lambda$); another is caused by the interaction between percentilization ($\mathbf{D}_Q$) and parameter shrinkage. Correction of these biases is complicated and requires the information of ratio of total abundance at adjacent time points ($\frac{c_t}{c_{t+1}}$) that is usually missing when using relative abundance data. Nonetheless, based on our results (Fig. S3), the estimated interaction strength remained positively correlated with the theoretical expectation, although the bias was not corrected.

**III. Parameterization of theoretical models**

We validated MDR S-map method using simulated time series data generated from the multi-species Ricker model,

$$N_i(t+1) = N_i(k)\exp\left[r_{0i}\left(1+e_i\mathbf{MN}(t)\right)\right] \text{------(Eq. 3.)},$$

where $r_{0i}$ is intrinsic growth rate of species $i$, $e_i$ is a vector taking zero values for all except for the $i$th entity, $\mathbf{M}$ represents time-independent interaction matrix with the size 1000 x 1000 ($m_{ij}$ represent the effect of species $j$ on species $i$). The species-specific intrinsic growth rate $r_{0i}$ was generated as

$$r_{0i} = r_{00}\left(1.0 + v_{r_0}U_1\right),$$

where $r_{00} = 1.5$ represents the average growth rate in the community, and $v_{r_0} = 0.5$ represents the size of inter-specific variation of the growth rate. Here, $U_1$ is a random variable following uniform distribution $Uniform(-1, 1)$. The interaction matrix $\mathbf{M} = (m_{ij})$, effect of $sp.j$ on $sp.i$ was generated as,

$$m_{ii} = \{-1 \quad 1 \leq i \leq n$$
$$m_{ij} \sim N_T(0, \sigma_I, I_{max}) \quad 1 \leq i \leq n-1, i+1 \leq j \leq n$$
$$m_{ij} = \begin{cases} m_{ji}(1+v_m U_1) & \text{if } m_{ji} < 0 \\ -m_{ji}(1+v_m U_1) & \text{else} \end{cases} \quad 2 \leq i \leq n,\ 1 \leq j \leq i-1$$

, where $N_T$ is a random variable following truncated normal distribution $N_T(\mu=0, \sigma=\sigma_I, \varsigma=I_{max})$ with probability density function involving the truncation threshold $\varsigma$,

$$f_{N_T}(N_T \mid \mu, \sigma, \varsigma) = \frac{g(N_T)}{F_N(\varsigma) - F_N(-\varsigma)}, \quad \begin{cases} g(N_T) = f_N(N_T; \mu, \sigma) & \text{if } -\varsigma \leq N_T \leq \varsigma \\ g(N_T) = 0 & \text{else} \end{cases}$$

In this function, $f_N$ and $F_N$ are the probability density function and cumulative distribution function of normal distribution $Normal(\mu, \sigma)$, respectively. Here, we let the mean interaction strength $\mu=0$, the inter-specific variation of interaction strength $\sigma_I = 2.0$, and the maximum interaction strength $I_{max} = 0.5$ (i.e., truncation threshold $\varsigma$). In addition, $v_m = 0.9$ determines the maximum difference of the interaction strength between $m_{ij}$ and $m_{ji}$. With this definition, the interaction matrix includes competition ($m_{ij} < 0$ & $m_{ji} < 0$) and trophic-interactions between resource sp.$j$ and consumer sp.$i$ ($m_{ij} > 0$ & $m_{ji} < 0$) only. Also note that the trophic interaction can occur only when $i < j$. Finally, we generated initial population size as $N_i(t_0) = n_0(1 + N_T)$, where $n_0$=0.1 and $N_T$ is a truncated normal random variable with $\mu$=0, $\sigma$=0.2, and $\varsigma$=0.9.

**IV. Testing robustness of the MDR S-map against data noise**

*IV-1. Random noises generation*

To test the robustness of MDR S-map method to data noise, we considered three types of random noises, including: i) observation noise, ii) process noise, and iii) stochastic environmental forcing. Firstly, for observation noise, we added noises into the (true) population size at each time step $t$. The inclusion of noises is carried out *after* the generation of time series data based on Eq. 3.,

$$n_i(t_k) = N_i(t_k)(1 + E_O U_1) \quad 1 \leq j \leq n, \ \forall t_k \in T : \{t_1, t_2, \ldots, t_N\},$$

where $E_O$ represents the size of observation noise. This is the conversion from the (true) population size vector $N(t)$ into the observed vector $n(t) = (n_1(t), n_2(t), \ldots, n_{1000}(t))^T$. For process noises and stochastic environmental forcing, we modeled the intrinsic growth rate of each species $i$ ($r_{i0}$ in Eq. 3) as a function of time-dependent process noises (from unknown mechanisms) $\varepsilon_i(t)$ and environmental forcing $env(t)$ as follows,

$$r_i(t) = r_{i0}\left[1 + E_E env(t) + E_P \varepsilon_i(t)\right],$$

where $r_{i0}$, $E_E$, and $E_P$ is the constant intrinsic growth rate of species $i$, species-independent effect size of environmental forcing, and species-independent effect size of process noise, respectively. In addition, $env(t)$ and $\varepsilon_i(t)$ are time-dependent environmental forcing with first-order autocorrelation (i.e., red noise) and time- and species-dependent noises without autocorrelation (i.e., white noise), respectively. Unlike observation noise, both process noises and stochastic environmental forcing were added in the iterative computation of difference equations (Eq. 3). Specially, the process noises were generated as $\varepsilon_i(t) = U_1$ and the environmental forcing $env(t)$ was generated as

$$env(1) = 0.1,$$
$$env(t) = \rho_{env} env(t-1) + (1-\rho_{env}) v_{env} U_1 \quad t \geq 2,$$

where $\rho_{env} = 0.8$ represents the degree of autocorrelation and $v_{env} = 0.5$ represents the maximum noise size. Based on various magnitude of $E_E$, $E_P$, and $E_O$, we generated the observed vectors **n** using identical initial values and parameters $\mathbf{N}(0)$, **M**, and $r_{i0}(k)$. In total, we considered four levels for each noise type, including zero, low, medium, and high level of noise ($E_E = E_P = E_O = 0.05, 0.1,$ and $0.15$ for low, medium, and high level, respectively). Then, we analyzed the time series data generated from the different levels of noise combinations (i.e., $4^3 = 64$ combinations). Results for testing influences of data stochasticity on MDR S-map were summarized in Figs. S4-7.

*IV-2. Robustness of MDR S-map to random noises*

The estimated interaction strengths by MDR S-map were robust against low to moderate levels of noise, including observation noises, process noises and stochastic environmental forcing (Fig. S4). Among various types of noises, the MDR S-map was relatively more sensitive to observation noises than the other types of noises; nevertheless, reasonable positive associations between estimated and theoretical interaction strengths were still maintained, even under moderate observation noise. Interestingly, however, the negative influences of observational noises were relieved when process noises exist (Fig. S5), although process noises undermined the performance of one-step forward forecast (Fig. S6). These analyses indicate that process noises, while bringing negative impacts on one-step forward forecast, facilitate quantification of interaction strengths. This finding, although seemingly counter-intuitive, can be explained with dynamical system theory; facilitative effects from process noises might help to better recover the local geometry of attractor when time series data are limited and sparse under high-dimensional settings. That is, process noises randomly travel data points to the states that otherwise will not be visited if observations are limited, and thus relieves data sparsity. Indeed, the positive influences of process noises become less visible when increasing observations (doubling the length of time series) (Fig. S7). Finally, our method was also robust to unseen environmental forcing (here, modeled as red noise), indicating that the estimated interaction strengths are still reliable even if some important external forces were not included in the analysis for practical reasons.

## V. Importance of embedding dimension in the estimation of network topology

In this study, we emphasize that an important feature of MDR S-map was its capacity of accommodating a large number of network nodes while still operating at the optimal embedding dimension through multiview ensemble approach. This was critically different from the approach used in a recent study (Ushio 2020) that quantified interactions by embedding all causal nodes into regularized S-map (Cenci *et al.* 2019) (i.e. the embedding dimension increases with the number of causal nodes) instead of operating at the optimal embedding dimension, an issue known as *over-embedding* (Kennel *et al.* 1992). Over-embedding uses a large number of coordinates, which makes SSR operations using finite time series data vulnerable to noises (Liebert *et al.* 1991). This is because some embedded spaces containing no or very limited dynamical information are included in SSR (Habeeb *et al.* 2005). Consequently, over-embedding regularized S-map makes the estimated topological properties deviated substantially from theoretical expectations. In contrast, the topological properties estimated by MDR S-map are more reliable and robust to various types of random noises (Figs. S9-10). Indeed, MDR S-amp accesses better quality of reconstructed interaction strengths than that accessed from over-embedding regularized S-map (Fig. S11), albeit that the over-embedding method roughly quantified individual interaction strengths in a relative sense. Therefore, accurate estimations of entire network properties (i.e., network topology and stability measures) remains difficult by the over-embedding approach, especially when the number of data points is not large compared to the embedding dimensions, as shown in our model examples. We noted that sample size may not be the issue in this particular study (Ushio 2020), because this study considered a network with very low linking density and thus the embedding dimensions were generally low compared to sample size (i.e., most $E < 20$ and $N = 122 \times 5 = 610$ time points used in the study (Ushio 2020)); however, it remains an open question concerning how many data points are needed for the over-embedding approach to operate. In summary, our analysis revealed the strong negative impacts of high-dimensionality on network reconstructions, especially for network topology, if the SSR was not operated at the optimal embedding dimension. As explained in previous sections, this was because precise neighborhood measures (e.g., distance) required for SSR-based approaches (Sugihara *et al.* 1990; Kennel *et al.* 1992; Shalizi 2006) cannot be obtained from a high-dimensional dataset (Bellman 1957), but can only be determined at the optimal (usually low) embedding dimension. Therefore, inferring high-dimensional networks through appropriate measures in SSR, e.g., multiview distance, was critical to determine correct interaction strength and topology of large networks.

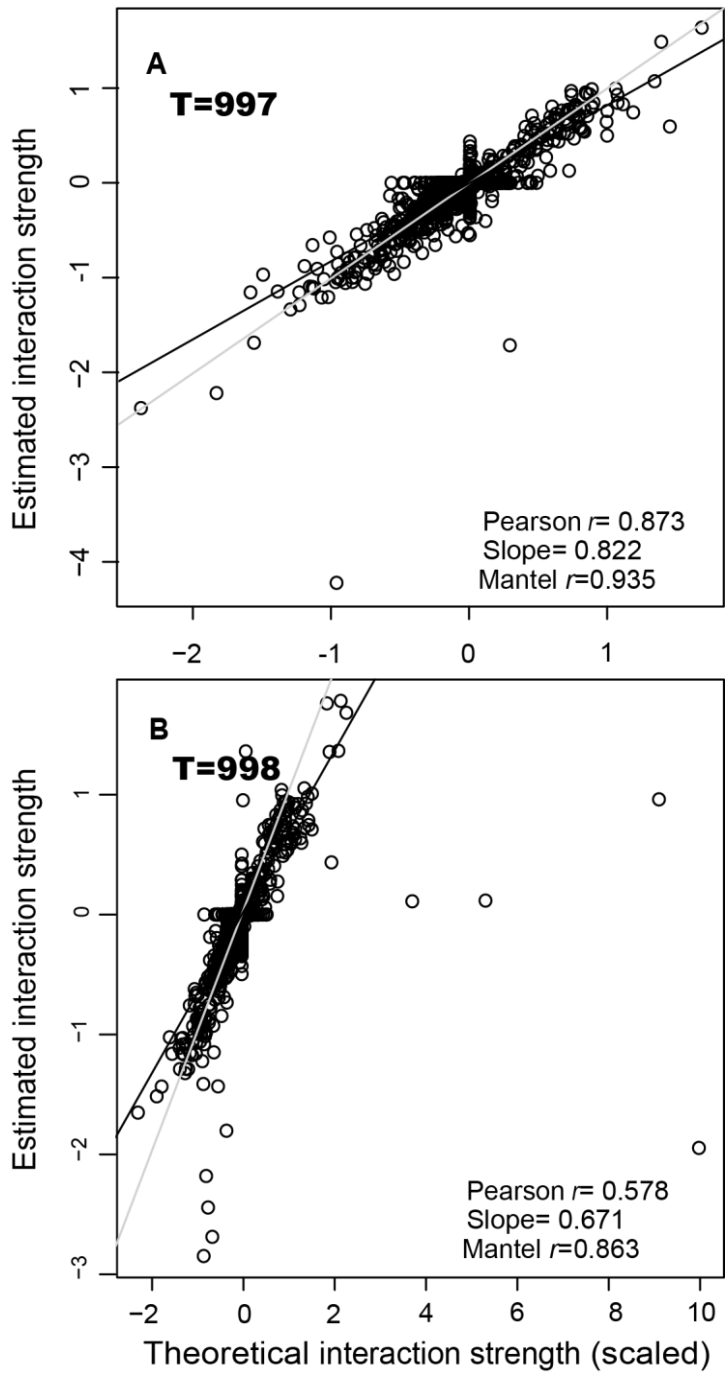

**Fig. S1. Examples illustrating that interaction strengths estimated for out-of-samples observed at T=997 and 998 were consistent with theoretical expectations.** Based on our analysis, the one-step-forward forecast can not only be used to predict future values of network nodes (Table 1), but also to reconstruct the whole interaction networks at future time points. Here, the grey line represents the 1:1 lines, and black lines represents the slope of quantile regression based on median. In Panel (*B*), most of the points are distributed along the 1:1 line; however, a few outliers are influential to the estimation of regression slope, which make the regression line deviate from the grey line and decrease Pearson *r*. In summary, once we know

the state of network nodes at previous time points, we can incorporate this information to the nonparametric MDR S-map model established by library time series and forecast the interaction networks for one or a few steps forward.

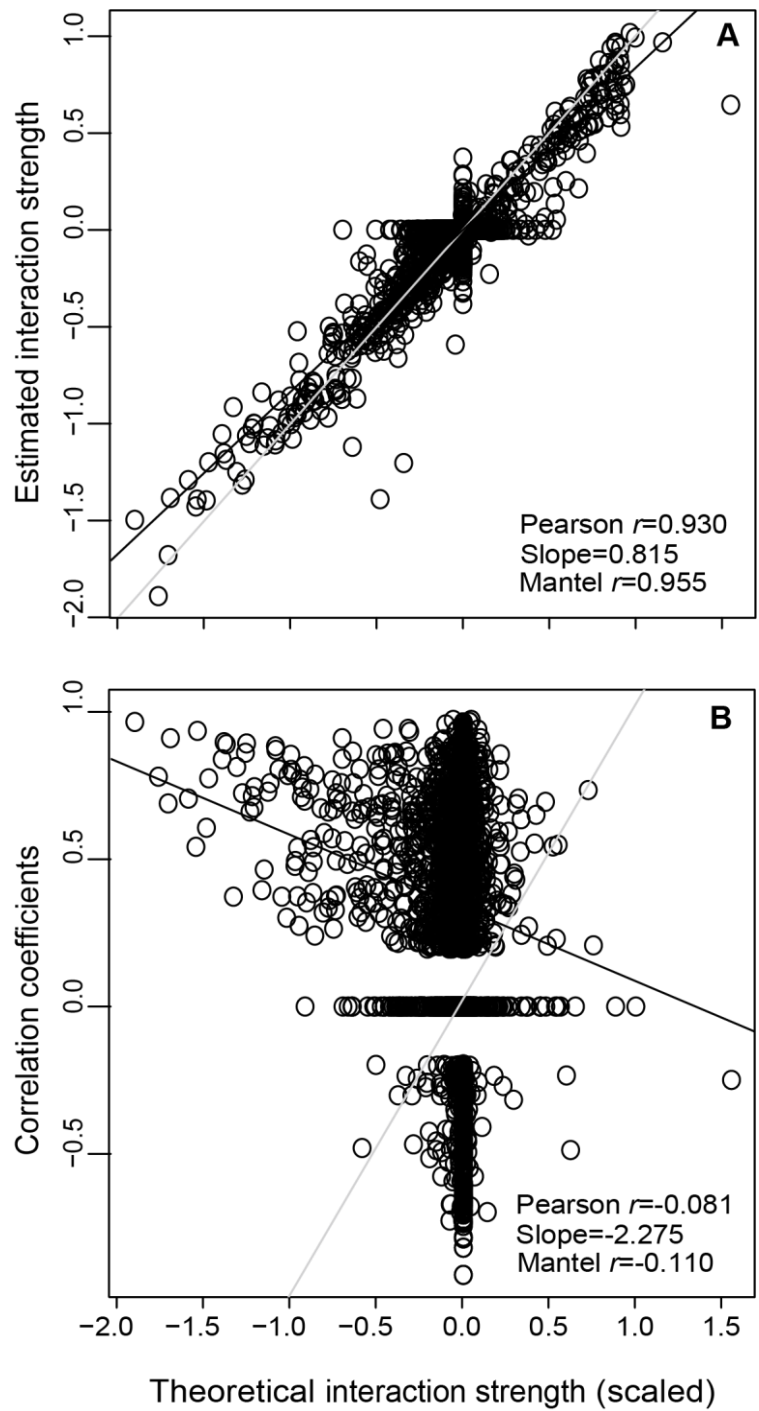

Fig. S2. **Long-term median of theoretical expectations was consistent with interaction strength estimated by MDR S-map, but not with the interaction strength inferred from correlation analyses.** Long-term median of interaction (i.e., Jacobian) matrices was calculated as the temporal median of each interaction strength observed at all time points. (*A*) Since the results for each time point had a strong positive relationship between MDR S-map estimations

and theoretical expectations (Fig. 2), the estimated long-term medians of interaction strengths corresponded well to the theoretical medians. However, (*B*) these theoretical medians had a very weak, even negative relationship with the correlation coefficients obtained from time series pairs. Here, the grey line represents the 1:1 lines, whereas the black line represents the slope of quantile regression based on median.

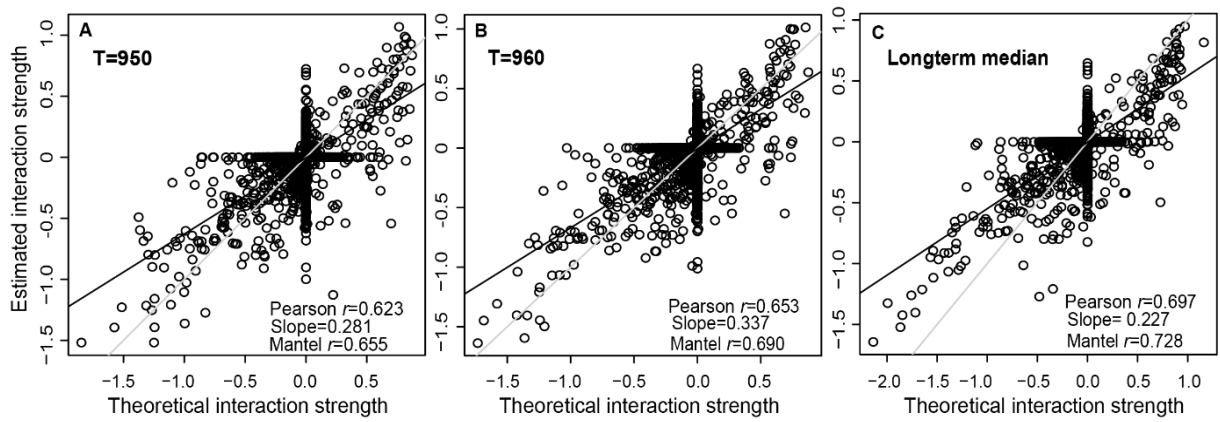

**Fig. S3. Interaction strengths estimated based on percentage (or relative abundance) data.** Estimations of interaction networks using MDR S-map had good correspondence with theoretical expectations for each time point, T=950 and 960 as examples (*A-B*) and their long-term median (*C*). However, these relationships were not as strong as estimations based on absolute abundance data, as percentilization introduced biases caused by unequally weighted data at adjacent time points (see details in SI Text, *II. Statistical properties of MDR S-map*). Here, the grey line represented the 1:1 lines and black line represented the slope of quantile regression, based on the median.

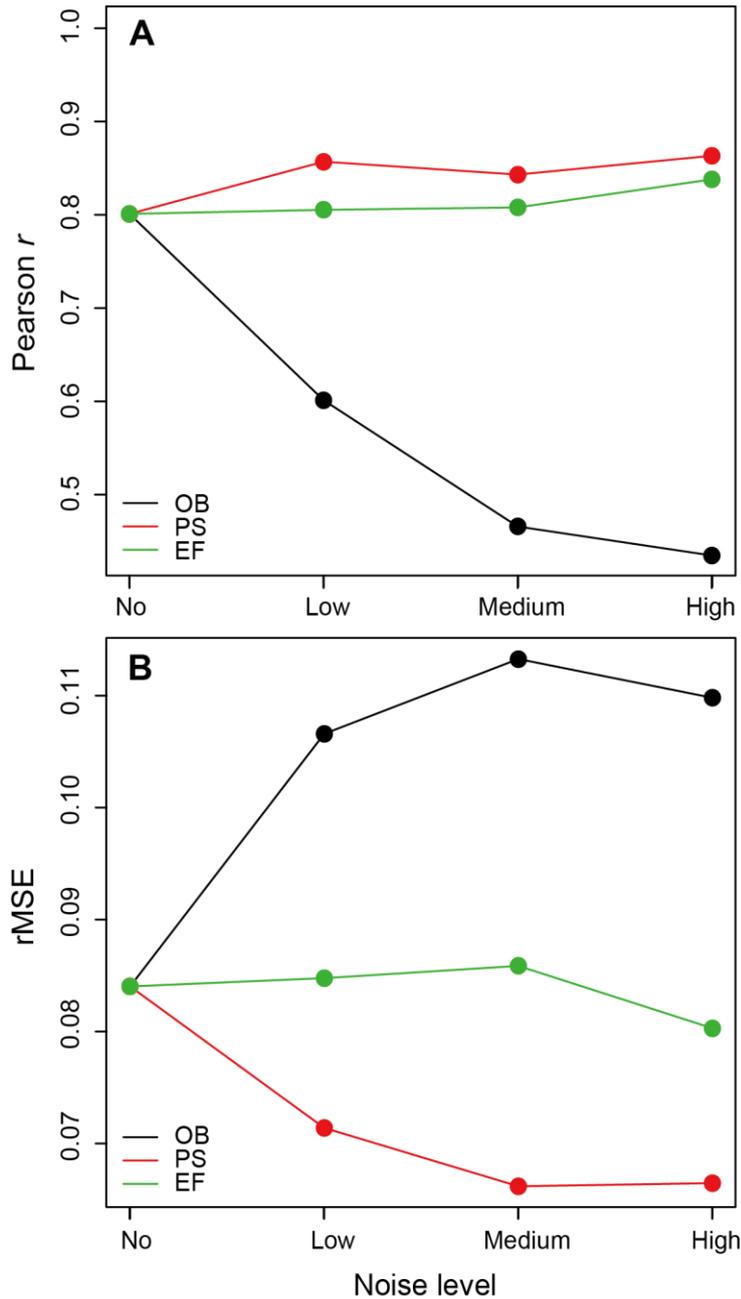

**Fig. S4. Interaction strengths estimated by MDR S-map are robust to different types of noises.** The prediction skills of estimated interaction strength were evaluated by calculating the temporal medians of (*A*) Pearson *r* and (*B*) rMSE between estimated and theoretical interaction strengths. Prediction skills varied with different types of noises, including observation noises (OB), process noises (PS), and stochastic environmental forcing (EF) (see details in SI Text, *IV. Testing robustness of the MDR S-map against data noise*). In this analysis, each type of noise was independently added (i.e. without the other types of noises). Among these types of noises, observation noises had the strongest impacts on prediction skills. Nonetheless, the positive relationships between the estimations and theoretical expectations remained robust, even under high levels of observation noises.

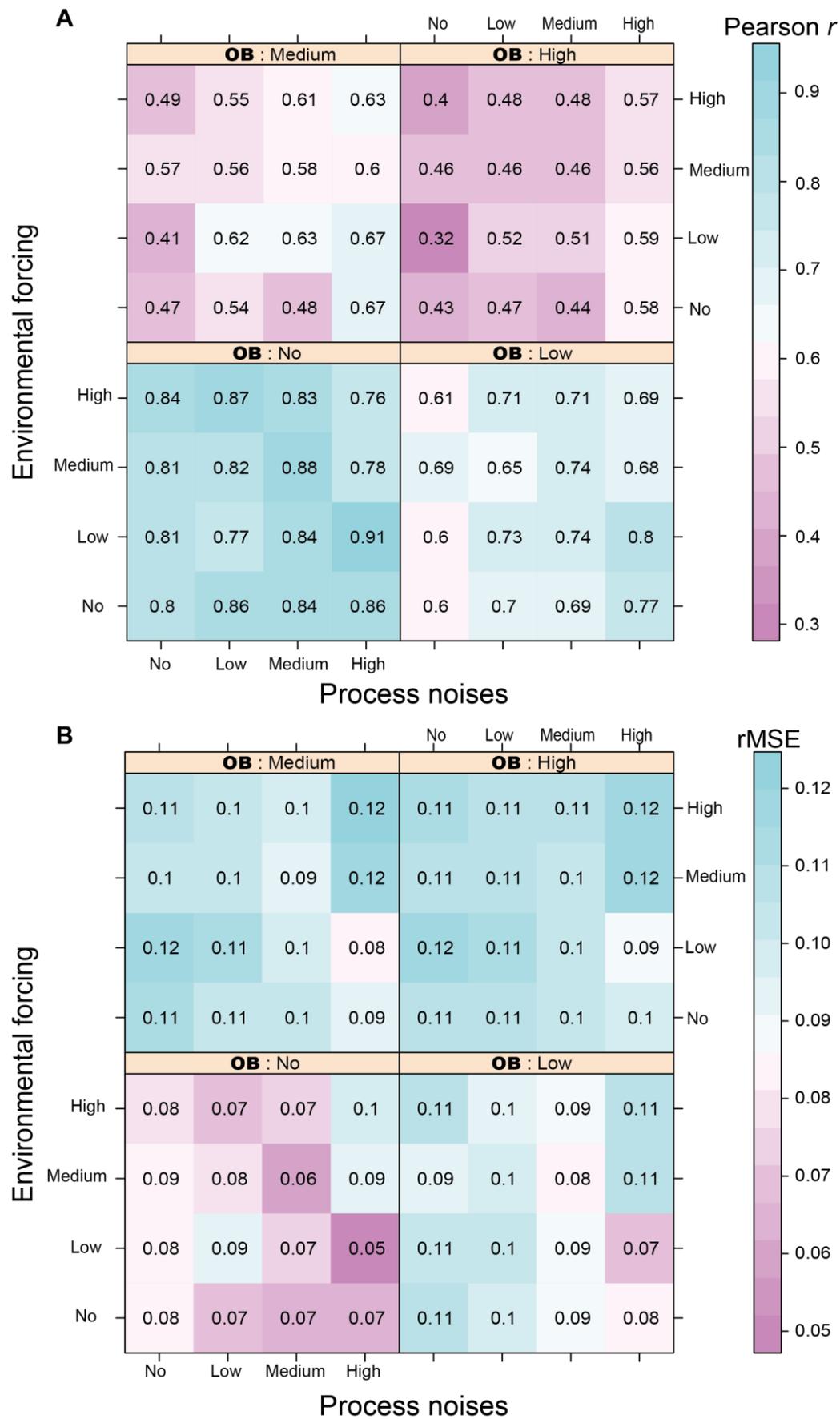

**Fig S5. Interaction strengths estimated by MDR S-map are robust to synergistic effects**

**of various types of noises**. Prediction skills of estimated interaction strength were evaluated by calculating the temporal medians of (*A*) Pearson *r* and (*B*) rMSE between the estimated and theoretical interaction strengths (see details in SI Text, *IV. Testing robustness of the MDR S-map against data noise*). These prediction skills consistently declined when elevating observation noises. However, both process noises and stochastic environmental forcing (modeled as red noises) buffer against the negative impacts of observation noises. Therefore, the estimations of interaction strengths reached the highest skills under strong process noises, especially when observation noises exist. The presence of process noises is mostly beneficial to our estimations, in particular when observation noises are strong. Similarly, although the presence of unseen environmental forcing also has positive influences on prediction skills, its influence is usually minor and sometimes even becomes negative when there are strong process noises.

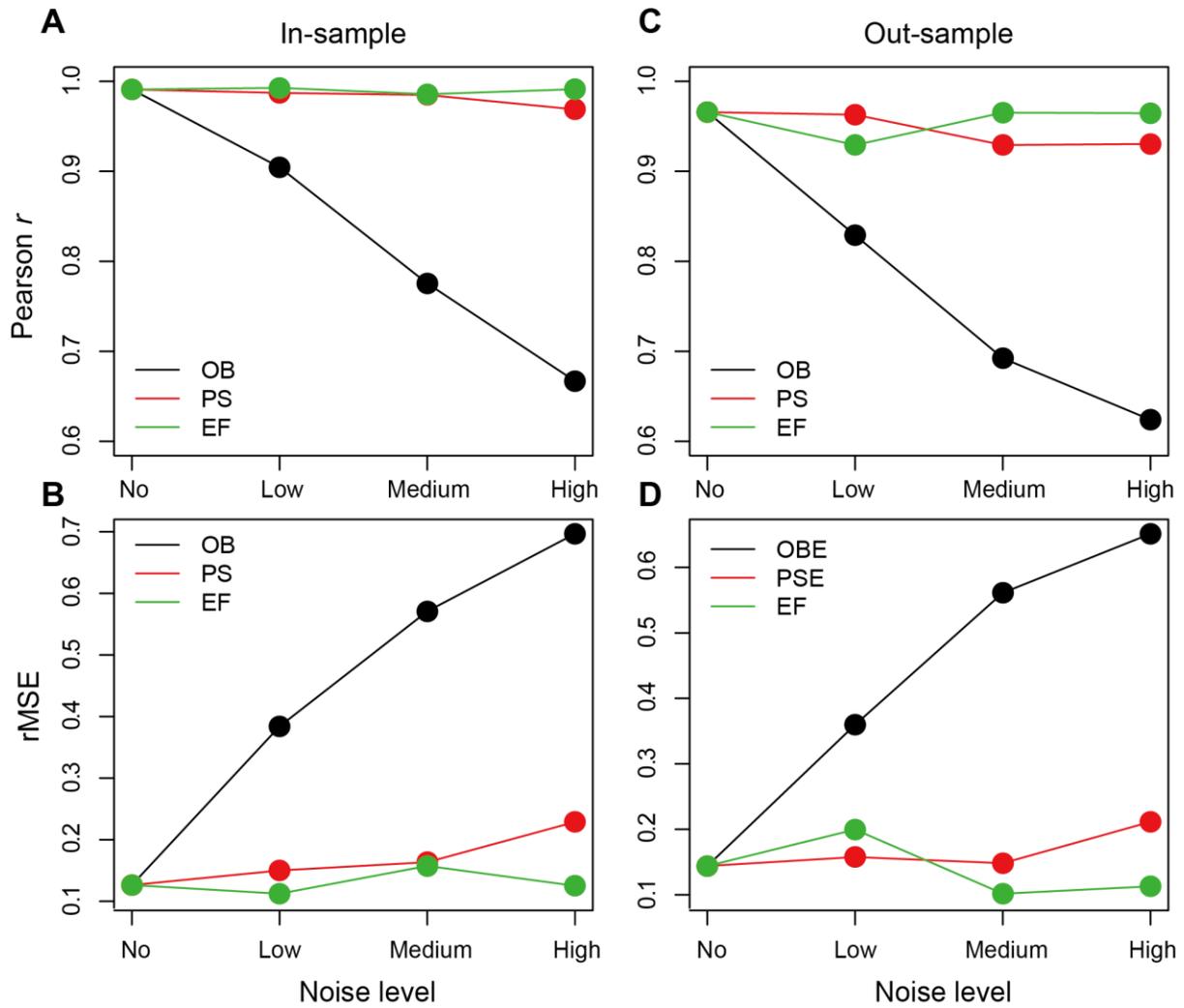

**Fig. S6. One-step forward forecast skills for noisy data.** The one-step forecast skills evaluated by (*A*, *C*) Pearson *r* and (*B*, *D*) rMSE between the predictions of MDR S-map and time series data (similar to the forecast skills shown in Table 1) varies with different types of noises, including observation noises (OB), process noises (PS), and stochastic environmental forcing (EF) which were independently added (i.e. without other types of noises) (see details in SI Text, *IV. Testing robustness of the MDR S-map against data noise*). Although estimations of interaction strengths improved when involving process noises (Fig. S4), performance of one-step forward forecast declined when elevating process noises in both leave-one-out cross-validation (In-sample, *A-B*) and out-of-sample forecast (Out-of-sample, *C-D*).

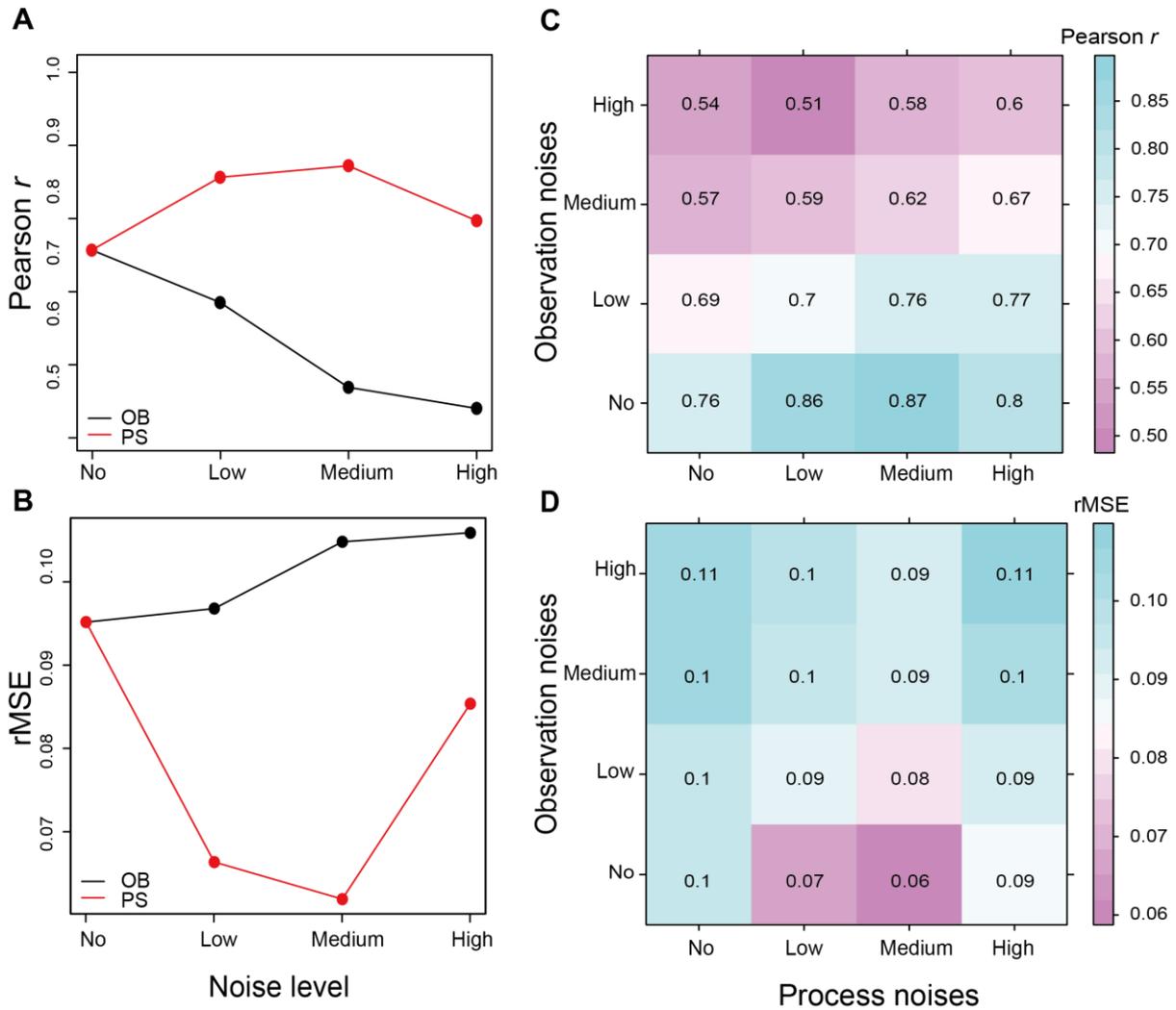

**Fig. S7. The robustness of MDR S-map to random noises using longer time series (n=200).** The prediction skills of estimated interaction strength, (*A*) Pearson *r* and (*B*) rMSE, were calculated to evaluated performance of MDR S-map using longer time series (*n*=200). Here, various types of noises, including observation noises (OB) and process noises (PS) were included (see details in SI Text, *IV. Testing robustness of the MDR S-map against data noise*). Similar to the analysis based on the shorter time series used in main texts (*n*=100), low to medium levels of process error still improved the prediction skill for inferring interaction strengths in the absence of observation error. However, strong process noises do not have positive influences on the estimations of interaction strength; this was in contrast to the positive effects of strong process noises observed in the analyses using shorter time series. In addition, for the synergistic effects of various types of noises (*C* and *D* are the prediction skills using Pearson *r* and rMSE, respectively), process noises buffered against the negative impacts of observation error, as observed in the analysis of short time series. However, this contrasted with the strong process error leading to lower prediction skill than adding a medium level of process error in the absence of observation error.

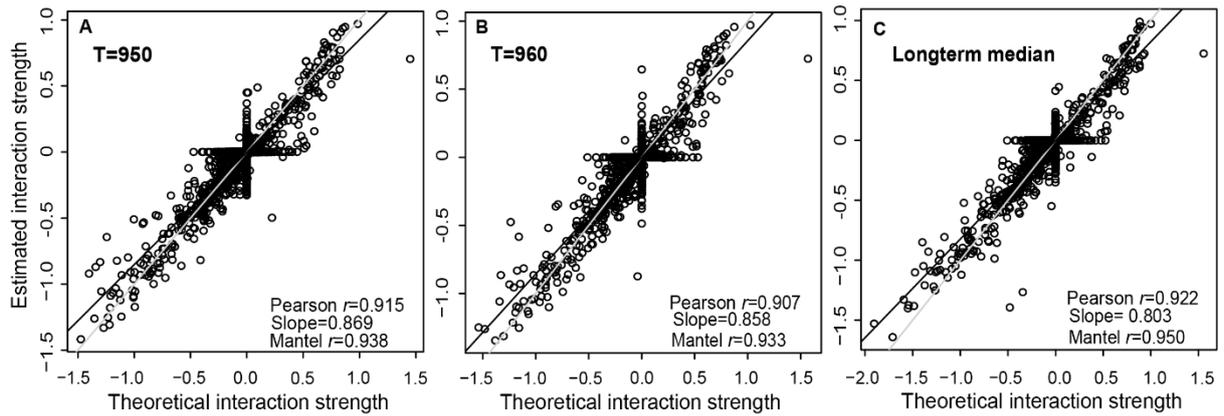

**Fig. S8. Interaction strengths estimated from incomplete network nodes.** Removal of some critical species will not influence estimations of interaction strengths among the remaining species. Here, dominant species with the top 10% relative abudnace (12 species out of 117 species) were removed from our analysis. However, removal of these critical species had very minor effects on the estimated interaction strengths among the remaining species (i.e., sub-networks). Compared to networks reconstructed using complete nodes, there was minor reduction in Pearson *r*, approximately 0.010, 0.013 and 0.09 for the networks reconstructed at (*A*) T=950, (*B*) T=960 and (*C*) the longterm medians (Fig. S2), respectively. That is, reconstruction of sub-network was still reliable, even if some network nodes or edges were not involved in the MDR S-map analysis for practical reseaons. Here, the grey line represents the 1:1 line, and black line represents the slope of quantile regression based on median.

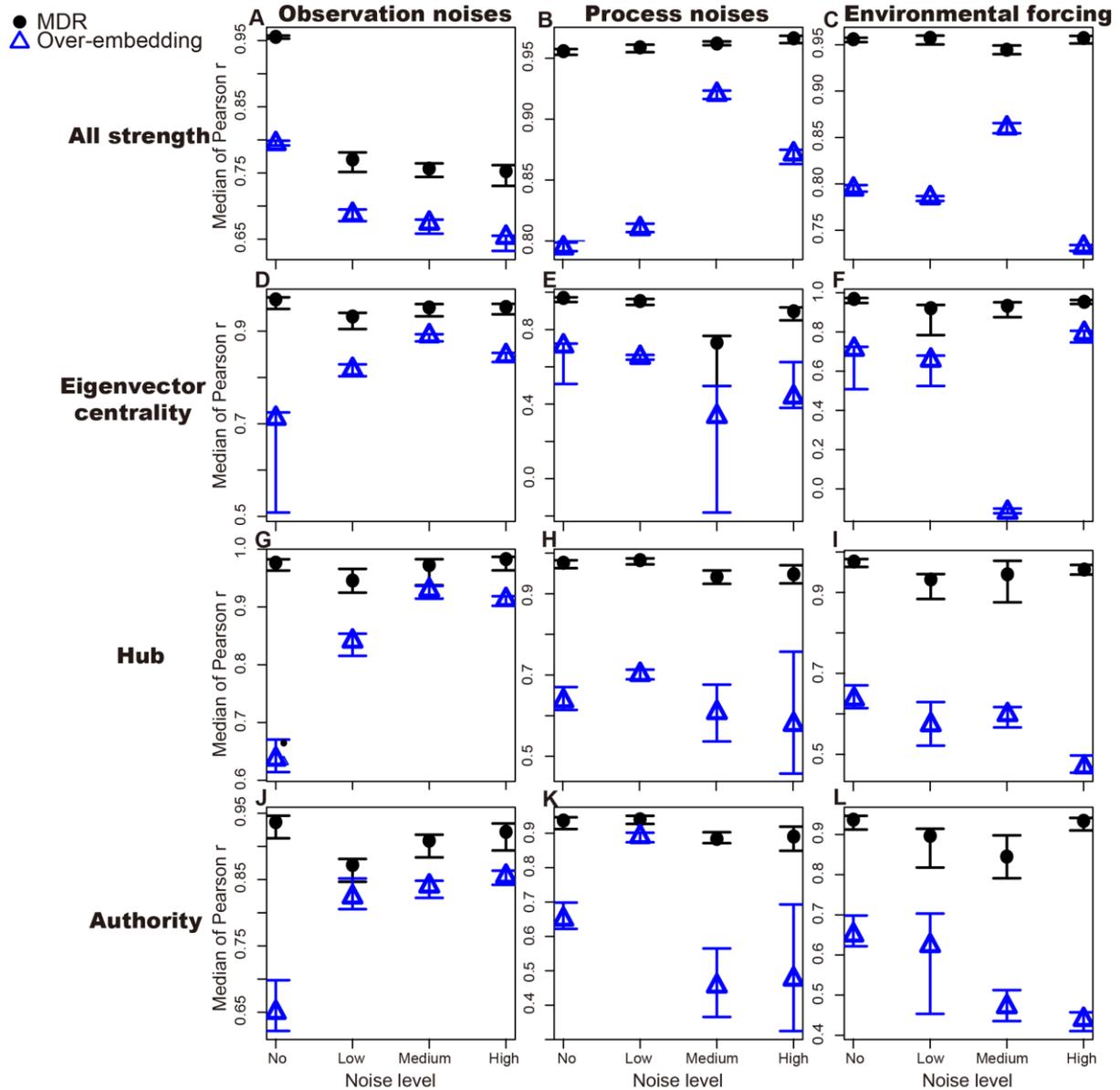

**Fig. S9. MDR S-map is robust to random noises and outperforms over-embedding regularized S-map in the estimations of node centrality indices.** The four centrality indices, all strength (i.e., in- + out- strength; *A-C*), eigenvector centrality (*D-F*), hub (*G-I*), and authority (**j-l**), were calculated for each node. Calculated indices changed with various types of noises, including observation noises (OB; *A*, *D*, *G*, and *J*), process noises (PS; *B*, *E*, *H*, and *K*), and unseen environmental forcing (EF; *C*, *F*, *I*, and *L*). To evaluate the performance of MDR S-map (black solid curcle) and over-embedding regularized S-map (blue open triangle; See details in SI Text, *V. Importance of embedding dimension in the estimation of network topology*), we calculated the median of time-varying correlation coefficients that measured correlations between theoretical and estimated nodes centrality at each time point. The error bars present 95% bootstrapped confidence intervals. We concluded that the estimation of node centrality based on MDR-S-map was robust to random noises. Throughout all noise scenarios, MDR S-map outperformed over-embeddeding regularized S-map in the estimation of node centrality.

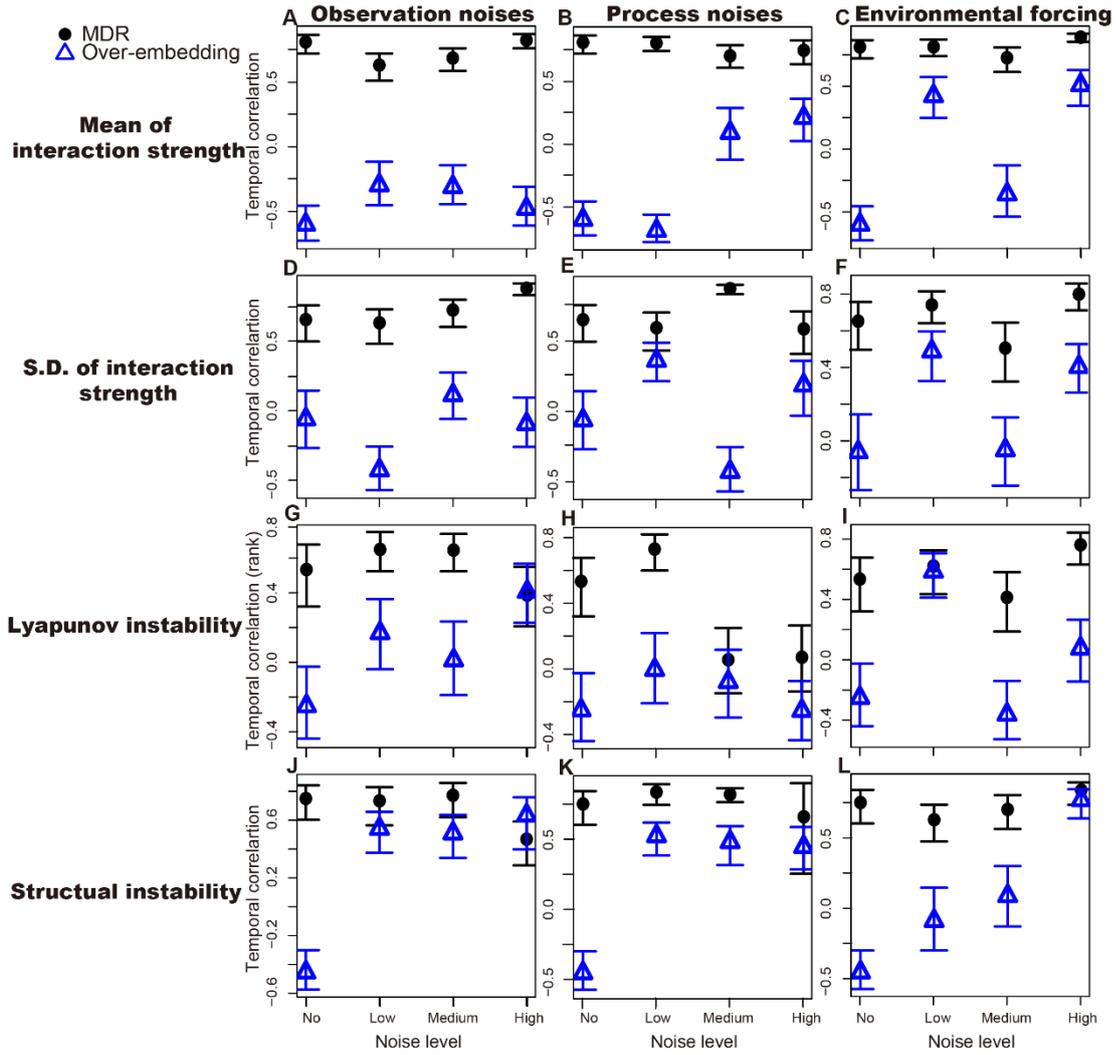

**Fig. S10. MDR S-map was robust to random noises and outperformed over-embedding regularized S-map in the estimations of network topology and stability indices.** Network topology, including (*A-C*) mean and (*D-F*) standard deviation of interaction strength and network stability indices, including (*G-I*) Lyapunov instability (*J-L*) and structual instability, were calculated from time-varying interaciton matrix. The calculated indices changed with various types of noises, including observation noises (OB; *A*, *D*, *G*, and *J*), process noises (PS; *B*, *E*, *H*, and *K*), and unseen environmental forcing (EF; *C*, *F*, *I*, and *L*). To evaluate the performance of MDR S-map (black solid circle; See details in SI Text, *V. Importance of embedding dimension in the estimation of network topology*) and over-embedding regularized S-map (blue open triangle), we calculated temporal correlations among these time-varying indices. The error bars represent 95% bootstrapped confidence intervals. The estimated network topology and stability based on MDR-S-map was robust to random noises. Throughout all noise scenarios, MDR S-map outperformed over-embeddeding regularized S-map in estimation of network topology. For estimation of network stability, the MDR S-map has greater performance in most scenarios, except for a few scenarios where an over-emebedding S-map performed similarly.

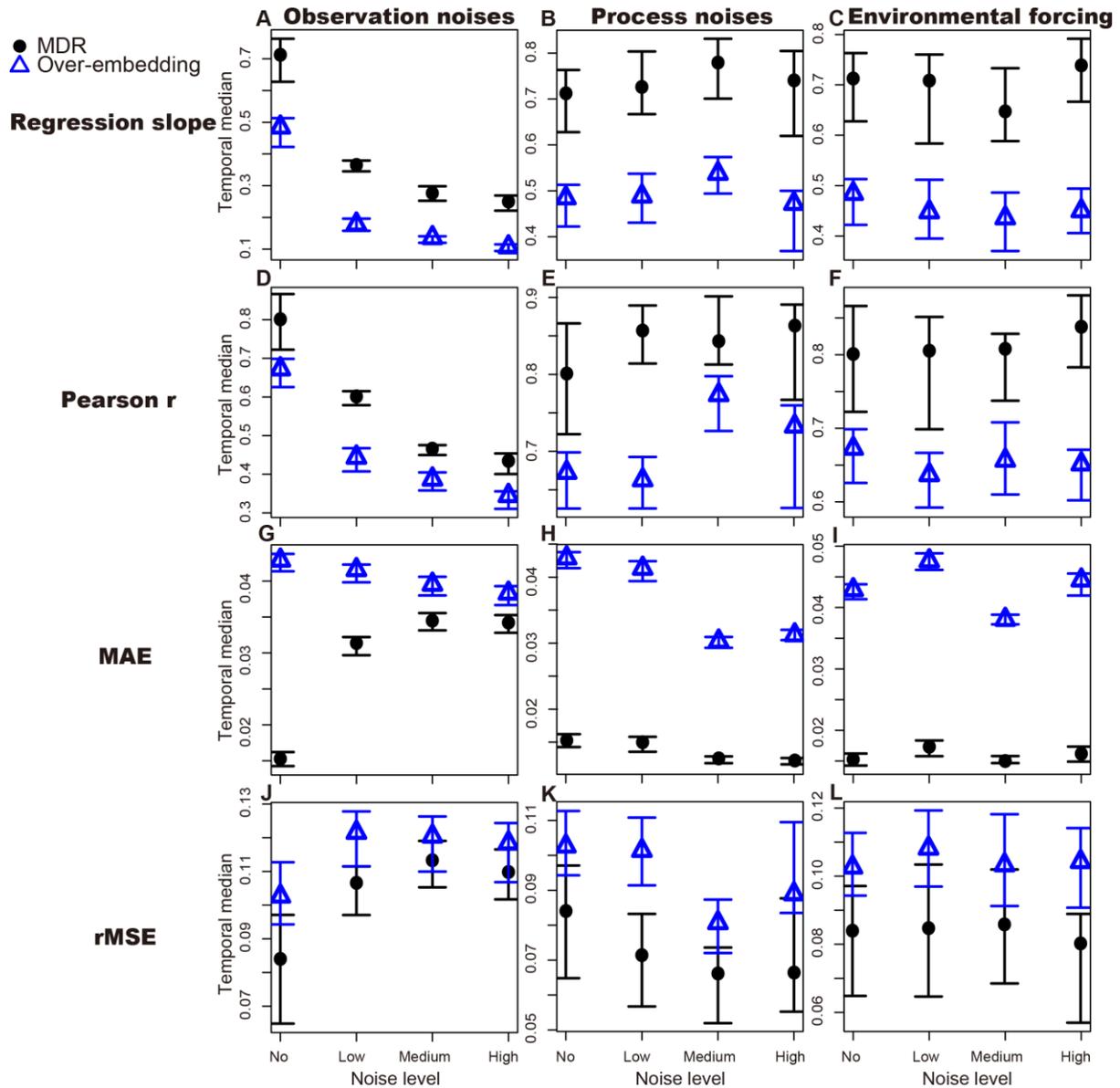

**Fig. S11. MDR S-map was robust to random noises and outperformed over-embedding regularized S-map in estimation of individual interaction strengths.** Here, three types of noises, including observation noises (OB; *A, D, G,* and *J*), process noises (PS; *B, E, H,* and *K*), and unseen environmental forcing (EF; *C, F, I,* and *L*), were considered as presented in Fig. S4. To evaluate method performance of MDR S-map (black solid curcle) and over-embedding regularized S-map (blue open triangle; See details in SI Text, *V. Importance of embedding dimension in the estimation of network topology*), we calculated the temporal median of the time-varying skills, including (*A-C*) regression slope, (*D-F*) Pearson correlation, (*G-I*) mean absoulte error (MAE), and (*J-L*) root mean square error (rMSE) computed from the compairson between estimated and theoretical interaction strengths at each time point. The error bars represent 95% bootstrapped confidence interval. The estimation of interaction strength based on MDR-S-map was robust to random noises. Throughout all noise scenarios, MDR S-map siginicantly outperformed over-embededing regularized S-map (non-overlapped confidence

intervals for regression slope, Pearson *r*, MAE, and some rMSE), although differences in Pearson correlation were usually not large (< 0.2 throughout the scenarios). We concluded that over-embedding regularized S-map roughly quantified the interaction strength, although problems caused by high dimensionality were not fully addressed.

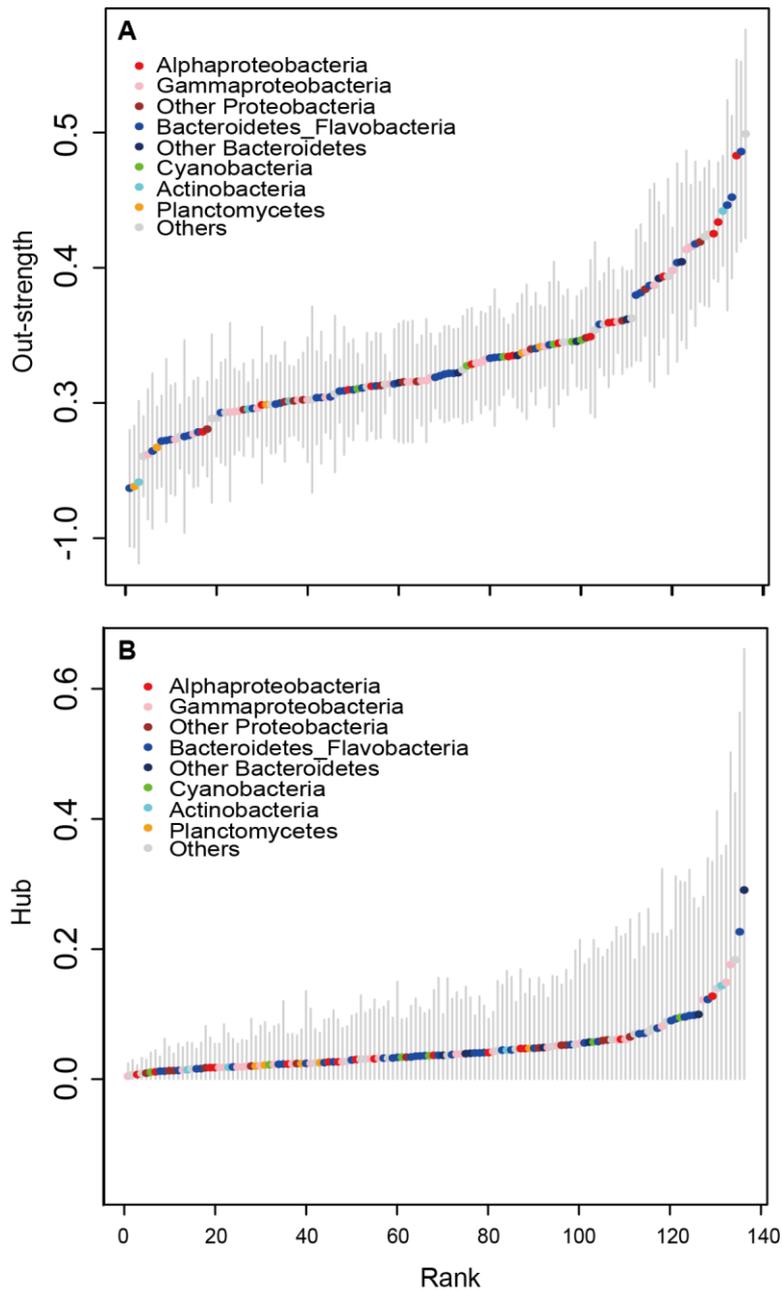

**Fig. S12. Importance of bacteria OTUs inferred from the reconstructed interaction networks.** Importance of each of the 137 bacteria OTUs was evaluated by (*A*) strengths of outward interactions affecting the other species (i.e., out-strength) and (*B*) hub considering not only the interaction strengths of the node itself, but also authority of the nodes it affected. The points indicate species ranked by their temporal mean of the topological indices. Grey bars indicate the temporal standard deviation (± 1SD); colors of points indicate various taxonomical groups.

**Table S1. Comparison of the topological properties characterizing the network nodes in reconstructed versus theoretical networks.** For the interaction network reconstructed at every time point, we calculated the topological indices for each node and compared them to those calculated from theoretical networks. To evaluate the performance of estimations, we computed the Pearson *r* and the slope of median-based regression by comparing estimations and theoretical expectations. The interaction network can be evaluated at each time point; for demonstration purposes, we summarized these indices in this table (mean ± SD; median). In particular, we applied a Kolmogorov-Smirnov test (K-S test) to test whether the shape of the cumulative distribution of estimated strengths (in-, out-, and all) differed from the distribution of theoretical expectations. There was rarely significant difference between estimations and expectations, except for the distribution of out-strengths.

| Type | In-strength | Out-strength | All strength | Eigenvector centrality | Hub | Authority |
|---|---|---|---|---|---|---|
| Pearson *r* | 0.90 ± .05; 0.90 | 0.97 ± .01; 0.97 | 0.96 ± .02; 0.96 | 0.76 ± .38; 0.96 | 0.83 ± .22; 0.94 | 0.77 ± .27; 0.91 |
| Slope | 0.87 ± .05; 0.87 | 0.89 ± .07; 0.88 | 0.85 ± .09; 0.84 | 0.85 ± .45; 1.00 | 0.69 ± .34; 0.84 | 0.75 ± .39; 0.83 |
| *p*-value of K-S test | 0.77 ± .19; 0.79 | 0.06 ± .11; 0.004 | 0.36 ± .35; 0.17 | -- | -- | -- |

16  **Table S2. Averaged node centrality index of dominant bacterial OTU ranked by mean out-strength.** The importance of OTU may be different
17  when concerning different centrality indices. Therefore, we labeled the top 10 highest mean values with respect to each centrality index by red
18  background.

| OTU | Out-strength | In-strength | All strength | Hub | Authority | Eigenvector centrality | Phylum | Class | Order | Family | Genus |
|---|---|---|---|---|---|---|---|---|---|---|---|
| 1 | 1.99 | 0.07 | 2.06 | 0.14 | 0.01 | 0.06 | Unidentifiable | | | | |
| 2 | 1.86 | 0.32 | 2.18 | 0.04 | 0.04 | 0.06 | Bacteroidetes | Flavobacteria | Flavobacteriales | Flavobacteriaceae | |
| 3 | 1.83 | 0.56 | 2.39 | 0.03 | 0.05 | 0.10 | Proteobacteria | Alphaproteobacteria | Rickettsiales | SAR11 | Pelagibacter |
| 4 | 1.52 | -0.05 | 1.48 | 0.10 | 0.01 | 0.05 | Bacteroidetes | Flavobacteria | Flavobacteriales | Flavobacteriaceae | |
| 5 | 1.46 | 0.05 | 1.52 | 0.23 | 0.02 | 0.21 | Bacteroidetes | Flavobacteria | Flavobacteriales | Flavobacteriaceae | |
| 6 | 1.42 | 0.15 | 1.57 | 0.14 | 0.01 | 0.12 | Actinobacteria | Actinobacteria | Actinomycetales | Microbacteriaceae | |
| 7 | 1.34 | -0.01 | 1.33 | 0.13 | 0.05 | 0.15 | Proteobacteria | Alphaproteobacteria | Rhodobacterales | Rhodobacteraceae | |
| 8 | 1.25 | 0.30 | 1.55 | 0.02 | 0.01 | 0.04 | Proteobacteria | Alphaproteobacteria | Rhodobacterales | Rhodobacteraceae | Loktanella |
| 9 | 1.25 | 0.69 | 1.94 | 0.06 | 0.05 | 0.08 | Firmicutes | Clostridia | Clostridiales | | |
| 10 | 1.23 | 0.33 | 1.56 | 0.08 | 0.06 | 0.14 | Unidentifiable | | | | |
| 11 | 1.19 | 0.49 | 1.68 | 0.07 | 0.01 | 0.07 | Proteobacteria | Epsilonproteobacteria | | | |
| 12 | 1.18 | 0.10 | 1.28 | 0.10 | 0.01 | 0.06 | Bacteroidetes | Flavobacteria | Flavobacteriales | Cryomorphaceae | Crocinitomix |
| 13 | 1.15 | 0.13 | 1.28 | 0.06 | 0.00 | 0.05 | Proteobacteria | Gammaproteobacteria | Vibrionales | Vibrionaceae | Vibrio |
| 14 | 1.14 | 0.08 | 1.21 | 0.18 | 0.03 | 0.11 | Proteobacteria | Gammaproteobacteria | | | |
| 15 | 1.04 | 1.16 | 2.21 | 0.29 | 0.02 | 0.21 | Bacteroidetes | Sphingobacteria | Sphingobacteriales | | |
| 16 | 1.04 | 0.06 | 1.10 | 0.09 | 0.02 | 0.09 | Bacteroidetes | Flavobacteria | Flavobacteriales | | |
| 17 | 0.98 | 0.05 | 1.03 | 0.04 | 0.05 | 0.09 | Proteobacteria | Gammaproteobacteria | Oceanospirillales | Halomonadaceae | Cobetia |
| 18 | 0.94 | 0.32 | 1.26 | 0.04 | 0.05 | 0.08 | Unidentifiable | | | | |
| 19 | 0.94 | -0.21 | 0.72 | 0.01 | 0.06 | 0.08 | Proteobacteria | Alphaproteobacteria | Rhodobacterales | Rhodobacteraceae | Sulfitobacter |
| 20 | 0.92 | 0.12 | 1.05 | 0.10 | 0.02 | 0.12 | Bacteroidetes | | | | |
| 21 | 0.87 | 0.35 | 1.22 | 0.15 | 0.01 | 0.11 | Proteobacteria | Gammaproteobacteria | Pseudomonadales | Moraxellaceae | Psychrobacter |
| 22 | 0.87 | 0.01 | 0.88 | 0.04 | 0.00 | 0.04 | Bacteroidetes | Flavobacteria | Flavobacteriales | Flavobacteriaceae | |
| 23 | 0.84 | 0.11 | 0.95 | 0.01 | 0.01 | 0.04 | Proteobacteria | Epsilonproteobacteria | Campylobacterales | Campylobacteraceae | Arcobacter |
| 24 | 0.82 | 0.72 | 1.54 | 0.07 | 0.01 | 0.07 | Bacteroidetes | Flavobacteria | Flavobacteriales | Flavobacteriaceae | Tenacibaculum |
| 25 | 0.80 | 0.04 | 0.84 | 0.04 | 0.02 | 0.04 | Bacteroidetes | Flavobacteria | Flavobacteriales | Flavobacteriaceae | |
| 26 | 0.63 | 0.33 | 0.95 | 0.02 | 0.03 | 0.07 | Unidentifiable | | | | |
| 27 | 0.62 | 0.39 | 1.01 | 0.04 | 0.07 | 0.06 | Bacteroidetes | Sphingobacteria | Sphingobacteriales | Saprospiraceae | |
| 28 | 0.61 | 0.06 | 0.67 | 0.03 | 0.03 | 0.05 | Proteobacteria | | | | |
| 29 | 0.60 | 0.53 | 1.13 | 0.01 | 0.07 | 0.09 | Proteobacteria | Gammaproteobacteria | Methylococcales | Methylococcaceae | Methylocaldum |
| 30 | 0.60 | 1.04 | 1.64 | 0.04 | 0.15 | 0.20 | Proteobacteria | Alphaproteobacteria | Rhodobacterales | Rhodobacteraceae | |
| 31 | 0.59 | 0.12 | 0.71 | 0.03 | 0.05 | 0.07 | Proteobacteria | Alphaproteobacteria | Rhodospirillales | Rhodospirillaceae | |
| 32 | 0.58 | 0.06 | 0.64 | 0.05 | 0.03 | 0.08 | Unidentifiable | | | | |

| # | | | | | | | Phylum | Class | Order | Family | Genus |
|---|---|---|---|---|---|---|---|---|---|---|---|
| 33 | 0.58 | 0.54 | 1.12 | 0.04 | 0.07 | 0.13 | Bacteroidetes | Flavobacteria | Flavobacteriales | Flavobacteriaceae | Tenacibaculum |
| 34 | 0.54 | 0.47 | 1.01 | 0.07 | 0.02 | 0.07 | Verrucomicrobia | Verrucomicrobiae | Verrucomicrobiales | | |
| 35 | 0.49 | -0.04 | 0.45 | 0.06 | 0.01 | 0.04 | Proteobacteria | Alphaproteobacteria | Rhodobacterales | Rhodobacteraceae | |
| 36 | 0.48 | 0.53 | 1.01 | 0.02 | 0.02 | 0.04 | Proteobacteria | Epsilonproteobacteria | Campylobacterales | Helicobacteraceae | Sulfurovum |
| 37 | 0.47 | 0.42 | 0.89 | 0.03 | 0.05 | 0.06 | Cyanobacteria | Cyanobacteria | | Family VIII | GpVIII |
| 38 | 0.46 | 0.21 | 0.67 | 0.04 | 0.05 | 0.08 | Bacteroidetes | | | | |
| 39 | 0.45 | 0.82 | 1.28 | 0.02 | 0.01 | 0.03 | Cyanobacteria | Cyanobacteria | | Family II | GpIIa |
| 40 | 0.45 | 0.19 | 0.64 | 0.07 | 0.07 | 0.10 | Unidentifiable | | | | |
| 41 | 0.45 | 0.12 | 0.56 | 0.01 | 0.02 | 0.05 | Unidentifiable | | | | |
| 42 | 0.44 | 0.16 | 0.60 | 0.05 | 0.07 | 0.07 | Proteobacteria | Alphaproteobacteria | Rhodobacterales | Rhodobacteraceae | |
| 43 | 0.43 | 0.45 | 0.88 | 0.09 | 0.04 | 0.13 | Cyanobacteria | Cyanobacteria | | | |
| 44 | 0.43 | 0.44 | 0.87 | 0.04 | 0.04 | 0.09 | Bacteroidetes | Flavobacteria | Flavobacteriales | | |
| 45 | 0.42 | 0.19 | 0.61 | 0.02 | 0.08 | 0.08 | Proteobacteria | Gammaproteobacteria | | | |
| 46 | 0.42 | 0.22 | 0.64 | 0.02 | 0.00 | 0.03 | Planctomycetes | Planctomycetacia | Planctomycetales | Planctomycetaceae | Planctomyces |
| 47 | 0.40 | 0.38 | 0.78 | 0.02 | 0.04 | 0.04 | Bacteroidetes | Flavobacteria | Flavobacteriales | Flavobacteriaceae | |
| 48 | 0.40 | 0.78 | 1.18 | 0.01 | 0.04 | 0.04 | Proteobacteria | Epsilonproteobacteria | Campylobacterales | Campylobacteraceae | Arcobacter |
| 49 | 0.38 | 0.00 | 0.37 | 0.08 | 0.02 | 0.09 | Proteobacteria | Gammaproteobacteria | | | |
| 50 | 0.37 | 0.29 | 0.67 | 0.02 | 0.02 | 0.02 | Planctomycetes | Planctomycetacia | Planctomycetales | Planctomycetaceae | Planctomyces |
| 51 | 0.35 | 0.24 | 0.59 | 0.05 | 0.02 | 0.07 | Bacteroidetes | Sphingobacteria | Sphingobacteriales | Saprospiraceae | Lewinella |
| 52 | 0.35 | 0.66 | 1.01 | 0.03 | 0.06 | 0.08 | Proteobacteria | Alphaproteobacteria | Rickettsiales | SAR11 | Pelagibacter |
| 53 | 0.34 | 0.25 | 0.60 | 0.01 | 0.02 | 0.03 | Proteobacteria | Alphaproteobacteria | Rhodobacterales | Rhodobacteraceae | Loktanella |
| 54 | 0.34 | 0.42 | 0.77 | 0.04 | 0.02 | 0.05 | Cyanobacteria | Cyanobacteria | | Family II | GpIIa |
| 55 | 0.34 | -0.97 | -0.63 | 0.01 | 0.13 | 0.08 | Bacteroidetes | Flavobacteria | Flavobacteriales | Flavobacteriaceae | |
| 56 | 0.34 | 0.30 | 0.63 | 0.03 | 0.01 | 0.04 | Bacteroidetes | Flavobacteria | Flavobacteriales | Flavobacteriaceae | |
| 57 | 0.33 | 1.60 | 1.93 | 0.01 | 0.07 | 0.13 | Bacteroidetes | Flavobacteria | Flavobacteriales | Flavobacteriaceae | |
| 58 | 0.32 | 0.15 | 0.47 | 0.02 | 0.02 | 0.04 | Proteobacteria | Gammaproteobacteria | | | |
| 59 | 0.30 | 0.83 | 1.13 | 0.02 | 0.05 | 0.04 | Proteobacteria | Gammaproteobacteria | | | |
| 60 | 0.30 | 0.47 | 0.77 | 0.03 | 0.03 | 0.03 | Proteobacteria | Gammaproteobacteria | | | |
| 61 | 0.29 | 0.11 | 0.40 | 0.04 | 0.02 | 0.06 | Proteobacteria | Alphaproteobacteria | Rickettsiales | SAR11 | Pelagibacter |
| 62 | 0.28 | 0.09 | 0.37 | 0.06 | 0.05 | 0.06 | Cyanobacteria | Cyanobacteria | | | |
| 63 | 0.25 | 0.45 | 0.70 | 0.03 | 0.04 | 0.03 | Verrucomicrobia | Verrucomicrobiae | Verrucomicrobiales | Verrucomicrobiaceae | Roseibacillus |
| 64 | 0.22 | 0.29 | 0.51 | 0.04 | 0.03 | 0.06 | Bacteroidetes | Sphingobacteria | Sphingobacteriales | | |
| 65 | 0.22 | 0.11 | 0.33 | 0.06 | 0.02 | 0.05 | Bacteroidetes | Flavobacteria | Flavobacteriales | Flavobacteriaceae | |
| 66 | 0.22 | 0.41 | 0.63 | 0.02 | 0.04 | 0.05 | Bacteroidetes | Flavobacteria | Flavobacteriales | Flavobacteriaceae | |
| 67 | 0.21 | 0.14 | 0.35 | 0.02 | 0.03 | 0.03 | Bacteroidetes | Flavobacteria | Flavobacteriales | Flavobacteriaceae | Gilvibacter |
| 68 | 0.20 | 0.58 | 0.79 | 0.01 | 0.03 | 0.05 | Bacteroidetes | Flavobacteria | Flavobacteriales | Flavobacteriaceae | |
| 69 | 0.19 | -1.42 | -1.23 | 0.04 | 0.15 | 0.11 | Bacteroidetes | Flavobacteria | Flavobacteriales | | |
| 70 | 0.19 | 0.44 | 0.63 | 0.02 | 0.05 | 0.04 | Proteobacteria | Gammaproteobacteria | Thiotrichales | Piscirickettsiaceae | Methylophaga |
| 71 | 0.17 | 0.22 | 0.38 | 0.06 | 0.07 | 0.08 | Proteobacteria | Gammaproteobacteria | | | |

| | | | | | | | | | | |
|---|---|---|---|---|---|---|---|---|---|---|
| 72 | 0.17 | -0.02 | 0.14 | 0.04 | 0.02 | 0.05 | Proteobacteria | Gammaproteobacteria | | | |
| 73 | 0.16 | 0.25 | 0.40 | 0.01 | 0.05 | 0.04 | Proteobacteria | Deltaproteobacteria | Desulfobacterales | Desulfobulbaceae | |
| 74 | 0.16 | 0.39 | 0.55 | 0.12 | 0.01 | 0.11 | Proteobacteria | Gammaproteobacteria | | | |
| 75 | 0.16 | 0.35 | 0.51 | 0.05 | 0.03 | 0.07 | Proteobacteria | Gammaproteobacteria | | | |
| 76 | 0.16 | 0.68 | 0.84 | 0.05 | 0.11 | 0.11 | Proteobacteria | Epsilonproteobacteria | Campylobacterales | Helicobacteraceae | Sulfurovum |
| 77 | 0.15 | 0.19 | 0.34 | 0.06 | 0.01 | 0.10 | Bacteroidetes | | | | |
| 78 | 0.14 | 0.28 | 0.42 | 0.04 | 0.01 | 0.04 | Bacteroidetes | Flavobacteria | Flavobacteriales | Flavobacteriaceae | Krokinobacter |
| 79 | 0.14 | 0.01 | 0.15 | 0.00 | 0.00 | 0.01 | Proteobacteria | Gammaproteobacteria | | | |
| 80 | 0.14 | 0.22 | 0.36 | 0.01 | 0.01 | 0.02 | Verrucomicrobia | Verrucomicrobiae | Verrucomicrobiales | Verrucomicrobiaceae | Roseibacillus |
| 81 | 0.13 | 0.00 | 0.12 | 0.02 | 0.06 | 0.06 | Proteobacteria | | | | |
| 82 | 0.13 | 0.42 | 0.55 | 0.04 | 0.05 | 0.04 | Bacteroidetes | Flavobacteria | Flavobacteriales | Flavobacteriaceae | Winogradskyella |
| 83 | 0.12 | 0.03 | 0.16 | 0.02 | 0.03 | 0.03 | Proteobacteria | Alphaproteobacteria | Rhodobacterales | Rhodobacteraceae | |
| 84 | 0.12 | 0.79 | 0.91 | 0.03 | 0.05 | 0.08 | Unidentifiable | | | | |
| 85 | 0.11 | 0.09 | 0.20 | 0.02 | 0.00 | 0.02 | Bacteroidetes | Flavobacteria | Flavobacteriales | Flavobacteriaceae | |
| 86 | 0.11 | 0.84 | 0.95 | 0.01 | 0.01 | 0.03 | Cyanobacteria | Cyanobacteria | | Family II | GpIIa |
| 87 | 0.10 | 0.31 | 0.40 | 0.03 | 0.02 | 0.04 | Bacteroidetes | Flavobacteria | Flavobacteriales | Flavobacteriaceae | |
| 88 | 0.10 | -0.04 | 0.05 | 0.03 | 0.06 | 0.05 | Proteobacteria | Alphaproteobacteria | Rhodobacterales | Rhodobacteraceae | |
| 89 | 0.09 | 1.55 | 1.64 | 0.06 | 0.04 | 0.09 | Bacteroidetes | Flavobacteria | Flavobacteriales | Flavobacteriaceae | |
| 90 | 0.09 | 0.11 | 0.20 | 0.02 | 0.03 | 0.02 | Bacteroidetes | Flavobacteria | Flavobacteriales | Flavobacteriaceae | |
| 91 | 0.07 | 0.08 | 0.15 | 0.09 | 0.01 | 0.12 | Verrucomicrobia | Verrucomicrobiae | Verrucomicrobiales | | |
| 92 | 0.05 | 0.47 | 0.51 | 0.05 | 0.03 | 0.05 | Bacteroidetes | Flavobacteria | Flavobacteriales | Flavobacteriaceae | |
| 93 | 0.04 | 0.63 | 0.67 | 0.02 | 0.02 | 0.04 | Proteobacteria | Gammaproteobacteria | | | |
| 94 | 0.04 | 0.14 | 0.19 | 0.02 | 0.02 | 0.05 | Bacteroidetes | Flavobacteria | Flavobacteriales | Flavobacteriaceae | |
| 95 | 0.04 | 0.38 | 0.42 | 0.03 | 0.02 | 0.04 | Bacteroidetes | Flavobacteria | Flavobacteriales | Flavobacteriaceae | |
| 96 | 0.02 | 0.04 | 0.07 | 0.18 | 0.01 | 0.12 | Unidentifiable | | | | |
| 97 | 0.02 | 0.17 | 0.19 | 0.02 | 0.07 | 0.09 | Proteobacteria | Gammaproteobacteria | | | |
| 98 | 0.02 | 0.75 | 0.77 | 0.06 | 0.12 | 0.20 | Proteobacteria | Betaproteobacteria | Burkholderiales | Oxalobacteraceae | |
| 99 | 0.02 | 0.27 | 0.29 | 0.03 | 0.02 | 0.05 | Proteobacteria | Gammaproteobacteria | Alteromonadales | Colwelliaceae | Colwellia |
| 100 | 0.02 | 0.70 | 0.72 | 0.02 | 0.04 | 0.04 | Proteobacteria | Epsilonproteobacteria | Campylobacterales | Campylobacteraceae | Arcobacter |
| 101 | 0.01 | 0.54 | 0.55 | 0.02 | 0.01 | 0.03 | Actinobacteria | Actinobacteria | Actinomycetales | | |
| 102 | 0.01 | 0.48 | 0.49 | 0.06 | 0.00 | 0.06 | Proteobacteria | Deltaproteobacteria | Desulfobacterales | Desulfobulbaceae | Desulfotalea |
| 103 | 0.00 | 0.45 | 0.45 | 0.03 | 0.01 | 0.03 | Bacteroidetes | Flavobacteria | Flavobacteriales | Flavobacteriaceae | |
| 104 | -0.01 | 0.38 | 0.38 | 0.09 | 0.05 | 0.08 | Bacteroidetes | Flavobacteria | Flavobacteriales | Flavobacteriaceae | Lutimonas |
| 105 | -0.01 | 0.60 | 0.59 | 0.04 | 0.03 | 0.03 | Fusobacteria | Fusobacteria | Fusobacteriales | Fusobacteriaceae | |
| 106 | -0.01 | 0.06 | 0.05 | 0.02 | 0.07 | 0.08 | Planctomycetes | Planctomycetacia | Planctomycetales | Planctomycetaceae | Planctomyces |
| 107 | -0.01 | 0.06 | 0.04 | 0.05 | 0.01 | 0.04 | Proteobacteria | Alphaproteobacteria | Rhodobacterales | Rhodobacteraceae | |
| 108 | -0.04 | 0.48 | 0.44 | 0.02 | 0.04 | 0.04 | Proteobacteria | Gammaproteobacteria | Thiotrichales | Piscirickettsiaceae | Methylophaga |
| 109 | -0.04 | 0.19 | 0.15 | 0.04 | 0.00 | 0.03 | Bacteroidetes | Flavobacteria | Flavobacteriales | Flavobacteriaceae | |
| 110 | -0.05 | -0.28 | -0.33 | 0.01 | 0.03 | 0.03 | Actinobacteria | Actinobacteria | Actinomycetales | | |

| | | | | | | | | | | |
|---|---|---|---|---|---|---|---|---|---|---|
| 111 | -0.05 | 0.36 | 0.31 | 0.05 | 0.01 | 0.04 | Proteobacteria | | | |
| 112 | -0.06 | 0.44 | 0.38 | 0.02 | 0.01 | 0.04 | Proteobacteria | Gammaproteobacteria | | |
| 113 | -0.06 | 0.29 | 0.23 | 0.02 | 0.03 | 0.03 | Proteobacteria | Gammaproteobacteria | | |
| 114 | -0.07 | 0.14 | 0.08 | 0.05 | 0.01 | 0.05 | Proteobacteria | Gammaproteobacteria | | |
| 115 | -0.07 | 0.55 | 0.48 | 0.03 | 0.01 | 0.05 | Unidentifiable | | | |
| 116 | -0.07 | 0.31 | 0.24 | 0.03 | 0.05 | 0.10 | Bacteroidetes | Flavobacteria | Flavobacteriales | |
| 117 | -0.11 | 0.42 | 0.31 | 0.02 | 0.02 | 0.03 | Unidentifiable | | | |
| 118 | -0.11 | 0.27 | 0.16 | 0.05 | 0.03 | 0.05 | Unidentifiable | | | |
| 119 | -0.19 | 0.41 | 0.21 | 0.05 | 0.05 | 0.05 | Proteobacteria | Epsilonproteobacteria | Campylobacterales | Helicobacteraceae | Sulfurovum |
| 120 | -0.21 | 0.06 | -0.15 | 0.02 | 0.01 | 0.03 | Proteobacteria | Alphaproteobacteria | Rhodobacterales | Rhodobacteraceae | |
| 121 | -0.21 | 1.13 | 0.92 | 0.03 | 0.09 | 0.11 | Bacteroidetes | Flavobacteria | Flavobacteriales | Flavobacteriaceae | |
| 122 | -0.23 | 0.25 | 0.02 | 0.03 | 0.09 | 0.07 | Proteobacteria | Gammaproteobacteria | | | |
| 123 | -0.24 | 0.59 | 0.35 | 0.03 | 0.03 | 0.04 | Bacteroidetes | Flavobacteria | Flavobacteriales | Flavobacteriaceae | |
| 124 | -0.25 | 0.52 | 0.27 | 0.10 | 0.07 | 0.13 | Bacteroidetes | Flavobacteria | Flavobacteriales | Flavobacteriaceae | Tenacibaculum |
| 125 | -0.25 | 0.26 | 0.01 | 0.03 | 0.04 | 0.03 | Unidentifiable | | | |
| 126 | -0.27 | 0.32 | 0.05 | 0.04 | 0.06 | 0.06 | Proteobacteria | Gammaproteobacteria | | | |
| 127 | -0.27 | 0.86 | 0.59 | 0.05 | 0.02 | 0.06 | Bacteroidetes | Flavobacteria | Flavobacteriales | Flavobacteriaceae | |
| 128 | -0.28 | 0.37 | 0.09 | 0.12 | 0.01 | 0.09 | Bacteroidetes | Flavobacteria | Flavobacteriales | Flavobacteriaceae | Polaribacter |
| 129 | -0.28 | 0.30 | 0.02 | 0.07 | 0.05 | 0.08 | Bacteroidetes | Flavobacteria | Flavobacteriales | Flavobacteriaceae | Tenacibaculum |
| 130 | -0.33 | 0.19 | -0.14 | 0.03 | 0.07 | 0.08 | Planctomycetes | Planctomycetacia | Planctomycetales | Planctomycetaceae | Rhodopirellula |
| 131 | -0.35 | 0.51 | 0.15 | 0.04 | 0.06 | 0.08 | Bacteroidetes | Flavobacteria | Flavobacteriales | Flavobacteriaceae | |
| 132 | -0.38 | 0.55 | 0.17 | 0.05 | 0.02 | 0.04 | Proteobacteria | Gammaproteobacteria | | | |
| 133 | -0.39 | 0.01 | -0.38 | 0.05 | 0.03 | 0.04 | Verrucomicrobia | Verrucomicrobiae | Verrucomicrobiales | Verrucomicrobiaceae | Haloferula |
| 134 | -0.59 | 0.69 | 0.11 | 0.04 | 0.04 | 0.08 | Actinobacteria | Actinobacteria | | | |
| 135 | -0.62 | 0.28 | -0.34 | 0.05 | 0.01 | 0.03 | Planctomycetes | Planctomycetacia | Planctomycetales | Planctomycetaceae | |
| 136 | -0.63 | 0.79 | 0.15 | 0.08 | 0.07 | 0.08 | Bacteroidetes | Flavobacteria | Flavobacteriales | Flavobacteriaceae | Winogradskyella |

19
20

21  **Legends for Movies S1**
22  Time-varying interaction network of empirical bacterial communities sampled daily in Canoe
23  Beach, Boston (see Methods). For demonstration, we only included the reconstructed bacterial
24  interaction networks for the first 30 days.
25
26
27
28
29